\documentclass[12pt]{article}
\usepackage{amsmath} 
\usepackage{multirow} 
\usepackage{amsfonts}
\usepackage{amssymb,graphics,psfrag}
\usepackage{array,epsfig,multirow,stmaryrd,graphicx}
\usepackage{comment}
\usepackage{sectsty}
\def\hybrid{\topmargin -20pt    \oddsidemargin 0pt
        \headheight 0pt \headsep 0pt
        \textwidth 6.25in       
        \textheight 9 in       
        \marginparwidth .875in
        \parskip 5pt plus 1pt 
          \jot = 1.5ex
   }
\hybrid
\numberwithin{equation}{section}
\numberwithin{table}{section}

\sectionfont{\large\bfseries}
\subsectionfont{\normalsize\bfseries}
\subsubsectionfont{\normalsize\it}

\newcommand{\beq}{\begin{equation}}
\newcommand{\eeq}{\end{equation}}
\newcommand{\be}{\begin{equation}}
\newcommand{\ee}{\end{equation}}
\newcommand{\bea}{\begin{eqnarray}}
\newcommand{\eea}{\end{eqnarray}}   
\newcommand{\ben}{\begin{eqnarray*}}
\newcommand{\een}{\end{eqnarray*}}                  
\newcommand{\ba}{\begin{aligned}}
\newcommand{\ea}{\end{aligned}}
\newcommand{\bt}{\begin{tabular}}
\newcommand{\et}{\end{tabular}}
\newcommand{\bc}{\begin{center}}
\newcommand{\ec}{\end{center}}

\newcommand{\cO}{\mathcal{O}}

\newcommand{\cE}{\mathcal{E}}

\newcommand{\cC}{\mathcal{C}}
\newcommand{\cD}{\mathcal{D}}
\newcommand{\cL}{\mathcal{L}}

\newcommand{\cK}{\mathcal{K}}
\newcommand{\cN}{\mathcal{N}}
\newcommand{\cW}{\mathcal{W}}

\newcommand{\cA}{\mathcal{A}}

\newcommand{\cB}{\mathcal{B}}
\newcommand{\cF}{\mathcal{F}}

\newcommand{\cJ}{\mathcal{J}}
\newcommand{\cR}{\mathcal{R}}

\newcommand{\cV}{\mathcal{V}}

\newcommand{\cM}{\mathcal M}

\newcommand{\I}{\text{Im}}
\newcommand{\R}{\text{Re}}

\newcommand{\bj}{{\bar\jmath}}

\DeclareMathOperator{\vol}{vol}

\newcommand{\dd}{d}

\newcommand{\bbZ}{\mathbb{Z}}
\newcommand{\bbR}{\mathbb{R}}
\newcommand{\bbC}{\mathbb{C}}
\newcommand{\bbP}{\mathbb{P}}

\newcommand{\nn}{\nonumber}

\newcommand{\cref}{{\bf [check ref]}}

\newcommand{\simga}{\sigma}

\psfrag{n1}{$\nu_1$}
\psfrag{n2}{$\nu_2$}
\psfrag{n1'}{$\nu_1'$}
\psfrag{n2'}{$\nu_2'$}
\psfrag{n9}{$\nu_9$}
\psfrag{n10'}{$\nu_{10}'$}
\psfrag{t1}{$\tilde{\nu}_1$}
\psfrag{t2}{$\tilde{\nu}_2$}
\psfrag{t9}{$\tilde{\nu}_9$}
\psfrag{t1'}{$\tilde{\nu}_1'$}
\psfrag{t2'}{$\tilde{\nu}_2'$}
\psfrag{t10'}{$\tilde{\nu}_{10}'$}

\def\blfootnote{\xdef\@thefnmark{}\@footnotetext} 
\long\def\symbolfootnote[#1]#2{\begingroup%
\def\thefootnote{\fnsymbol{footnote}}\footnote[#1]{#2}\endgroup}

\begin{document}

\baselineskip=18pt

\begin{titlepage}
\begin{flushright}
\parbox[t]{1.8in}{
BONN-TH-2008-05\\
MAD-TH-2008-06\\
0805.3361\ [hep-th]}
\end{flushright}

\begin{center}

\vspace*{ 1.2cm}

{\large \bf  U(1) Mediation of Flux Supersymmetry Breaking}

\vskip 1.2cm

\begin{center}
 \bf{Thomas W.~Grimm and Albrecht Klemm} \footnote{grimm@th.physik.uni-bonn.de, aklemm@th.physik.uni-bonn.de}
\end{center}
\vskip .2cm

{\em Physikalisches Institut der Universit\"at Bonn, \\[.1cm]
Nussallee 12, 53115 Bonn, Germany}
\vspace*{.1cm}

and
\vspace*{.1cm}

{\em Department of Physics, University of Wisconsin, \\[.1cm]
        Madison, WI 53706, USA
        }

 \vspace*{1cm}

\end{center}

\vskip 0.2cm

\begin{center} {\bf ABSTRACT } \end{center}

We study the mediation of supersymmetry breaking triggered by background
fluxes in Type II string compactifications with $\cN=1$ supersymmetry.
The mediation arises due to an $U(1)$ vector multiplet coupling to both a hidden supersymmetry breaking 
flux sector and a visible D-brane sector. The required internal manifolds can be constructed 
by non-K\"ahler resolutions of singular Calabi-Yau manifolds.
The effective action encoding the $U(1)$ coupling 
is then determined in terms of the global topological properties of the internal space. 
We investigate suitable local geometries for the hidden and visible sector in 
detail. This includes a systematic study of orientifold symmetries of del
Pezzo surfaces realized in compact geometries after geometric transition. 
We construct compact examples admitting the key properties to realize flux
supersymmetry breaking and $U(1)$ mediation. Their toric realization allows us
to analyze the geometry of curve classes and confirm the topological connection between
the hidden and visible sector.

\hfill May, 2008
\end{titlepage}

\tableofcontents
\newpage

%
%

\section{Introduction}

The embedding of the standard model and its supersymmetric extensions 
into string theory is of crucial importance in the study of string phenomenology.
One promising arena for model building are Type II string compactifications 
with gauge theories realized on intersecting D-branes extending along
the four non-compact dimensions \cite{review_Dbrane}. Particularly appealing are scenarios 
for which the visible gauge theory and matter interactions can be localized within 
the internal space. Some crucial parts of the four-dimensional physics, such as the 
gauge group and matter content, are then determined by the local 
geometry near the intersecting standard model branes. This opens 
the door for concrete model building within a well-defined local framework
\cite{review_Dbrane,Douglas:1996sw,singMSSM,Model_F}.

In addition to the strings confined to the standard model branes, there 
will be string states which propagate through 
the bulk and interact with hidden sectors which are separated from the 
visible branes within the internal space.
In particular, supersymmetry breaking occurring in a hidden sector  
can be mediated to the visible branes via such messengers.
There are essentially two categories for the mediation of supersymmetry
breaking. Firstly, it can be mediated via gauge theory degrees of freedom
which arise from open strings in Type II compactifications \cite{Giudice:1998bp,SusyBreviews}. Secondly, the
mediation can occur due to closed string modes with Planck mass suppressed
couplings  \cite{SusyBreviews}. The latter category includes the 
anomaly mediation scenarios arising from the gravitational sector~\cite{Randall:1998uk}.
In contrast to the universal anomaly mediation, the contributions of gauge and
Planck suppressed mediation typically depend on the geometry and distances 
probed by the messengers between the visible and hidden sector. Different geometric set-ups 
can lead to a domination of one or the other mediation mechanism.

In this paper we investigate a string theory embedding of 
a mediation mechanism involving a $U(1)$ vector multiplet. 
If this vector multiplet is coupling to both the visible as well as some hidden
supersymmetry breaking sector, it can serve as a messenger of the breaking.
The phenomenological aspects of such scenarios depend on the role of the
$U(1)$ in the visible sector. Concrete proposals use  
the possible existence of additional $U(1)$ vectors, known as $Z'$ gauge bosons, which couple to 
the Standard Model \cite{Z'physics}. Mediation of supersymmetry breaking 
involving such extra $U(1)$ multiplets has been studied in various
contexts in refs.~\cite{U(1)mediation,Langacker}. Another recent proposal 
is to identify the mediating $U(1)$  with the hypercharge 
of the MSSM \cite{Dermisek:2007qi}. Viable soft supersymmetry breaking terms 
are generated if the bino gains a mass from supersymmetry breaking in the
hidden sector in addition to the anomaly mediation contribution to all soft terms.
Set-ups with soft terms induces by $U(1)$ mediation or anomaly mediation
exhibits a number of very attractive phenomenological features such as 
natural suppression of CP- and flavor violation and a solution 
of the $\mu$ problem.

In this work we specify a concrete Type IIB string compactification
with the appropriate $U(1)$ couplings and
mechanism to break supersymmetry.
More precisely, we argue that supersymmetry breaking due to non-vanishing R-R and NS-NS 
background fluxes can be naturally mediated by $U(1)$ vector multiplets. 
The messengers arise as linear combinations of hidden and 
visible sector $U(1)_{\rm H}$ and $U(1)_{\rm V}$ vector multiplets. 
The coupling of the $U(1)_{\rm H}$ and $U(1)_{\rm V}$ is obtained from  
a gauging of a R-R scalar, similar to mechanism recently studied 
in ref.~\cite{Verlinde:2007qk} applying the idea of~\cite{Witten:1985bz}. 
However, in our set-ups the hidden $U(1)_{\rm H}$ 
gauge fields arise from the R-R four-form and pair in the underlying $\cN=2$ theory 
with the complex structure deformations of the internal manifold into supermultiplets.
Exactly these scalars obtain an F-term in a non-supersymmetric flux background  \cite{Gukov:1999ya},
and therefore render the $U(1)$ gauginos massive. 
We show that simple topological relations connecting the visible and hidden sector 
within the internal space will ensure the generation of  soft supersymmetry breaking 
terms for the standard model fields.
In particular, we discuss how candidate internal geometries 
are constructed as non-K\"ahler resolutions of a singular Calabi-Yau
manifolds.

Supersymmetry breaking by  background fluxes has been
investigated since the advent of flux compactification \cite{review_flux}.  
In particular, soft supersymmetry breaking terms induced by Planck suppressed 
moduli mediation have been first computed in refs.~\cite{soft_flux}. 
It thus has to be argued that the $U(1)$ mediation will lead to a visible 
effect on the low energy masses. 
In general, direct couplings can be suppressed if the hidden
supersymmetry breaking flux sector is separated, or rather `sequestered' \cite{Randall:1998uk},
from the standard model branes.  This will be the 
case for set-ups where supersymmetry is broken by fluxes near a deformed 
singularity hidden in a warped throat away from the visible sector \cite{Kachru:2007xp}. This does however 
still permit that the soft terms are corrected due to anomaly mediation, which yields 
to a mixture of two contributions as in the scenarios of refs.~\cite{Choi:2005ge,Dermisek:2007qi}.
Simple supersymmetry breaking flux backgrounds on a deformed Calabi-Yau
singularity have been constructed in refs.~\cite{ABSV}, and
argued to be large-$N$ dual to meta-stable systems of $D5$ and anti-$D5$ branes. 
The orientifold versions of these models can serve as a hidden sector in our
compactifications. Even though our analysis is more general and explicit,
specific flux vacua will realize the large $N$ dual of the $U(1)$ mediation
scenario of ref.~\cite{Verlinde:2007qk}.

In the construction of compact examples we need to ensure that 
both a visible brane sector as well as a hidden supersymmetry 
breaking sector can be realized. Geometrically this is achieved by 
picking Calabi-Yau manifolds with the appropriate singularities. The
singularities are then resolved or deformed to yield a smooth compact 
internal manifold permitting $U(1)$ mediation. In particular, this will lead us 
to the study of del Pezzo singularities and their resolutions. A del Pezzo surface is 
of real dimension $4$ and  has a sufficiently substructure to support intersecting branes inducing a spectrum 
and gauge group of a MSSM like model \cite{singMSSM}. We will focus on the 
del Pezzo surfaces which are obtained by blowing up $\bbP^2$ at $5,6,7$ or $8$ points.
The blow up process and the specification of 
appropriate orientifold projections will be described in detail. 

In order to find an explicit realization of the $U(1)$ scenario, we 
have to be able to check the topological connection between the hidden and
visible sector. We will argue that this is possible by analyzing the global
embedding of the del Pezzo surface into the compact space. The computation of
the BPS invariants for concrete torically realized examples reveals a decisive criterion to analyze 
which of the two-cycles in the del Pezzo surface are non-trivial in the compact space and 
connected to the hidden sector. Since the geometry of del Pezzo surfaces is
captured by Lie algebras, this criterion can be
reformulated in terms of the decomposition of Lie algebra representations.

This paper is organized as follows. In section \ref{generalU(1)} 
the general mechanism of $U(1)$ mediated supersymmetry 
breaking and our sting theory embedding is summarized. 
This includes in section \ref{rev_U(1)} a brief overview 
of the  phenomenological properties of the scenarios suggested 
in refs.~\cite{Langacker,Dermisek:2007qi}. 
In section \ref{Compact} we discuss the basics on constructing the 
necessary internal geometries as well as the corresponding 
$\cN=1$ four-dimensional effective theories. The relevant 
non-K\"ahler resolutions are introduced in section
\ref{nonKaehler_res}. The $\cN=1$ orientifold projection as well as
details on the effective action are presented in section \ref{nonKaehler_ori}.
The candidate visible sectors arise from del Pezzo surfaces as 
discussed in section \ref{D7_sector}, while the hidden sector flux geometry inducing 
the supersymmetry breaking are studied in section \ref{hiddenSusy}.

In the second part of this work we turn to explicit geometrical constructions of the
outlined set-up. The hidden geometry are orientifolds of $A_{n}$ singularities introduced  
in section~\ref{singularitieswithsmallresolutions}. 
More effort is devoted to the study of del Pezzo transitions in compact 
Calabi-Yau spaces. We analyze a large class of candidate internal manifolds in 
section~\ref{delPezzogeometries} and \ref{numericalchange}, and discuss their orientifold symmetries. 
A compact manifolds with associated orientifold projection  
admitting most of the desired properties to permit $U(1)$ mediation is
constructed in section \ref{eF1}. The toric construction of the compact
geometries as well as further explicit examples
are provided in appendices~\ref{torictransitions} and~\ref{mgn}.

%
%

\section{$U(1)$ Mediation of Supersymmetry Breaking \label{generalU(1)}}

In this section we first describe the general mechanism how a 
$U(1)$ vector multiplet coupling to both a hidden as well 
as a visible sector can mediate supersymmetry breaking. 
We briefly discuss in section
\ref{rev_U(1)} the phenomenological implications in case this $U(1)$ coupling 
is one of the dominant mediation mechanisms. The general idea how to realize
such a scenario within a flux compactification
of type IIB string theory will be presented in section \ref{U(1)_string}.  

\subsection{The mediation mechanism and its phenomenology \label{rev_U(1)}}

Let us consider a four-dimensional $\cN=1$ supersymmetric 
theory consisting of a visible MSSM-like 
sector with Lagrangian $\cL_{\rm visible}$ and a hidden sector $\cL_{\rm hidden}$.
We denote the bosonic components of the supermultiplets 
in the visible sector collectively by $Q$, while 
they are denoted by $\Phi$ for the hidden sector. Supersymmetry breaking 
is assumed to take place in the hidden sector. 
The $U(1)$ mediation of this breaking is possible if the field 
dependence of the effective low energy Lagrangian is of the form
\beq \label{Lagr_split}
  \cL_{\rm eff} = \cL_{\rm visible}(Q,A) + \cL_{\rm hidden} (\Phi,A)\ , 
\eeq
where $A$ is a $U(1)$ gauge boson in a vector multiplet  coupling to both the visible and 
hidden sector. Generically both the hidden and visible sector will contain fields charged 
under $A$. The holomorphic gauge kinetic coupling function of $A$ will be denoted 
by $f(\phi)$
and will depend on the hidden sector chiral multiplets collectively denoted by $\phi$.

Consider now supersymmetry breaking by a  non-vanishing F-term  in 
the hidden sector. The coupling of $A$ to the hidden sector can yield 
a significant contribution $\tilde M$ to the mass of the fermion $\lambda$ in the vector multiplet $(A,\lambda)$.
Recall that in an $\cN=1$ supersymmetric theory, the gauge kinetic term for the superfield 
$\cW$ containing $(A,\lambda)$ is given by
\beq
\label{Gauge_Kinetic_U(1)}
\cL_{\rm kin}(W)=\int \dd \theta^2 \tfrac{1}{4} f(\phi)\, \cW^\alpha \cW_{\alpha}
+ {c.c.} \ .
\eeq
The F-terms in the hidden sector are denoted by $F_I,\, F^I$ and take the form 
\beq
   F_{I} = D_{\phi^I} W\ ,\qquad \qquad F^{I} = e^{K/2} K^{I\bar J} \bar F_{\bar J}\ .
\eeq
Here $\phi^I$ are complex scalars in the chiral multiplets, $W$ is the holomorphic 
superpotential, and $K,K^{I\bar J}$ are the K\"ahler potential and inverse K\"ahler metric.
It follows from \eqref{Gauge_Kinetic_U(1)} that the fermion $\lambda$
in the multiplet of the vector $A$ acquires a mass
\beq
\label{Bare_Bino_mass}
\tilde M= F^{I} \partial_{\phi^I} \log( \R f)\ . 
\eeq 
The presence of the massive fermion $\lambda$ coupling to the visible sector
can have a profound impact on the observed supersymmetry breaking
phenomenology.

In a generic string compactification with non-vanishing F-terms also other
mediation mechanism can contribute to the soft parameters in the visible
sector. Clearly, it needs to be ensured that additional gauge
interactions between the two sectors are subdominant. More severely, gravity
mediation can contribute to the soft parameters of order $F_I/M_P$.  
In the string compactifications we will consider later, these contributions 
can be suppressed due to sequestering \cite{Kachru:2007xp}. The dominant
supergravity contribution then arises from anomaly mediation. In section \ref{HyperAnomaly}
a mixing of $U(1)$ mediation with anomaly mediation is illustrated for the
specific model proposed in ref.~\cite{Dermisek:2007qi}.

To illustrate the phenomenological implications 
of $U(1)$ mediation we will briefly review
two recently proposed scenarios. 
In the first scenario 
the vector $A$ is the hypercharge $U(1)_{\rm Y}$ of the MSSM 
\cite{Dermisek:2007qi}, while 
in the second scenario it corresponds to an additional $U(1)'$ under which 
all MSSM particles are charged \cite{Z'physics,U(1)mediation,Langacker}. 
This overview is also meant to highlight the generic features which are eventually
demanded from a string realization.

\subsubsection{Hypercharged anomaly mediation \label{HyperAnomaly}}
 
In reference~\cite{Dermisek:2007qi} it was proposed to identify $A$ with the hypercharge $U(1)_{\rm Y}$ 
of the MSSM. It was assumed that the 
only source of supersymmetry breaking is a non-vanishing hidden 
sector F-term in \eqref{Bare_Bino_mass} and the contributions due to anomaly 
mediation \cite{Randall:1998uk}.
The mass $\tilde M$ in \eqref{Bare_Bino_mass} 
will contribute to the bare bino mass. 
This is the decisive change of boundary conditions fed 
into the bino renormalization group equation running from the compactification
scale to low energies. 
The breaking scale of anomaly mediation is set by the gravitino
mass $m_{{3}/{2}}=e^{{K}/{2}}| W|$.
The gaugino masses $M_1,M_2,M_3$, the soft masses $m_i$ as well 
as the Yukawa couplings $A_{ijk}$ in the visible sector are then given by
\beq \label{mass_spectrum}
\begin{array}{rlrl}
M_1=&\displaystyle{\tilde M+\frac{b_1 g_1^2}{8 \pi^2 } m_{{3}/{2}}}\ , \qquad \qquad \qquad \qquad& 
M_a=&\displaystyle{\frac{b_a g_a^2}{8 \pi^2 } m_{{3}/{2}} , \quad a=2,3}\ ,\\ [3 mm] 
m_i^2=&\displaystyle{-\frac{1}{32 \pi^2 }\frac{\dd \gamma_i}{\dd \log \mu }}
m_{{3}/{2}}\ , &
A_{ijk}=&\displaystyle{-\frac{\gamma_i+\gamma_j+\gamma_k}{16 \pi^2 } m_{{3}/{2}}}\ ,\\ [3 mm] 
\end{array}
\eeq
where $b_a$ are the $\beta$-function coefficients, and $\gamma_i$ are the
anomalous dimensions of the matter fields. The $g_i$ are the three gauge couplings
of the MSSM.

The difference of the soft terms \eqref{mass_spectrum} to
the ones arising in general gauge mediation scenarios \cite{Giudice:1998bp}
is that the supersymmetry breaking in the hidden sector only contributes at leading order 
to the bino mass $M_1$. The other soft masses are identical to the ones obtained for the
anomaly mediation scenario.
It was shown in ref.~\cite{Dermisek:2007qi}, that only for certain values of
$\tilde M$ the renormalization group equation flow
yields an acceptable low energy spectrum,
\beq
\label{Def_alpha}
\tilde M\equiv \alpha\, m_{3/2} \ ,
\qquad \qquad 0.05  \lesssim |\alpha| \lesssim 0.25  \ ,
\eeq 
for $m_{3/2} \gtrsim 35$ TeV.
In an explicit string realization of this scenario the supersymmetry 
breaking mechanism in the hidden sector has to induce F-terms 
generating an $\tilde M$ satisfying the bound \eqref{Def_alpha}.

\subsubsection{Mediation by an additional $U(1)'$ \label{U(1)'model}}

In this section we briefly review a scenario where $U(1)$
mediation of supersymmetry breaking arises due to an additional 
$U(1)'$ factor extending the MSSM gauge group~\cite{Z'physics,U(1)mediation,Langacker}.
In references \cite{Langacker} it was
proposed to extend the MSSM by an additional $U(1)'$ gauge symmetry under
which all MSSM fields, as well as a new Standard Model singlet $S$ are charged. 
Here $S$ replaces the $\mu$ parameter of the MSSM. This extended MSSM needs to include a number 
of exotics with Yukawa couplings to $S$ in order to cancel anomalies~\cite{Langacker}.
From the point of view of string theory, the presence of additional $U(1)$
symmetries and exotics is rather generic and models such as the 
one presented in \cite{Langacker}
might therefore admit a natural embedding into a string compactification. 

In the scenario of \cite{Langacker} it was demanded that the $U(1)'$ gauge symmetry is 
not broken in the hidden sector, but rather in the visible sector 
through a vev of the additional field $S$. In contrast,  
supersymmetry breaking is assumed to take place 
in the hidden sector. The non-vanishing F-terms generate a mass
$\tilde M$ for the fermion in the $U(1)'$ vector
multiplet directly as in~\eqref{Bare_Bino_mass} or through loop corrections involving 
the supersymmetry breaking scalars. 
For a Lagrangian of the form \eqref{Lagr_split} there are no direct couplings
to the hidden sector and supersymmetry breaking is mediated by the
$U(1)'$. Since all chiral multiplets are charged under the $U(1)'$ the fermion
soft masses $m_i$ are generated already by a one loop correction. 
The gauginos do not directly couple to the $U(1)'$ and are thus only induced 
at the two loop level.  Explicitly, the gaugino and scalar soft masses  
take the form
\beq
  M_a \sim \frac{\tilde g^2 g_a^2}{(16 \pi^2)^2} \tilde M \ \log\Big(
  \frac{\Lambda_S}{\tilde M}\Big)\ , \qquad \qquad  
  m_i^2 \sim \frac{\tilde g^2 Q_i^2}{16 \pi^2} \tilde M^2 \ \log\Big(
  \frac{\Lambda_S}{\tilde M}\Big)\ ,
\eeq
where $\tilde g$ is the $U(1)'$ gauge coupling, $Q_i$ 
are the $U(1)'$ charges of the matter multiplets, and $\Lambda_S$ is the scale
of supersymmetry breaking. We refer  the reader to ref.~\cite{Langacker} for a
detailed analysis of the phenomenology of this model.

In this work we will study the necessary steps for embedding $U(1)$ 
models such as the ones of section \ref{HyperAnomaly} and \ref{U(1)'model} 
in a Type IIB flux compactification. This requires a precise specification of the 
supersymmetry breaking hidden sector and its $U(1)'$ 
coupling to the visible sector.

\subsection{An embedding into string theory \label{U(1)_string}}

\begin{figure}[!ht]
\begin{picture}(200,120)
\put(90,10){\includegraphics[height=3.5cm]{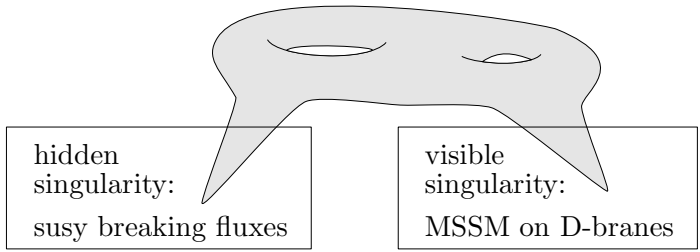}
} 
\put(105,48){\small hidden}
\put(105,37){\small singularity:}
\put(105,21){\small susy breaking fluxes}
\put(253,48){\small visible}
\put(253,37){\small singularity:}
\put(253,21){\small MSSM on D-branes}
\end{picture}
\vspace*{-.5cm} 
\caption{\small  \label{schematics} Compact manifold with hidden and visible
  sector singularity.}
\end{figure}

In the following we will describe a string theory scenario 
which admits a four-dimensional low energy effective Lagrangian of the 
form \eqref{Lagr_split}. 
We will consider a type IIB compactification on 
a manifold admitting two, possibly warped, singularities as schematically 
depicted in figure \ref{schematics}. The visible singularity is resolved by two- 
or four-dimensional cycles while the hidden singularity is deformed by three-cycles.
If the internal manifold obeys certain topological conditions,
we argue that such a set-up allows a $U(1)$ mediation of 
supersymmetry breaking triggered by background R-R and NS-NS three-form fluxes 
to a visible sector.

\subsubsection{Mixing hidden and visible $U(1)$ vector multiplets}

The string compactifications of interest admit
two $U(1)$ gauge fields $A_{\rm H}$ and $A_{\rm V}$ coupling 
to the hidden and the visible sector respectively. Their kinetic 
terms in an $\cN=1$ effective theory are of the form 
\beq
 \cL_{\rm kin}= \sum_{i= \rm V,H} \big( \tfrac12 \R f_i\, F_i \wedge * F_i + \tfrac12 \I f_i\, F_i \wedge
 F_i \big) \ ,
\eeq
where $F_{\rm V} = dA_{\rm V}$, $F_{\rm H} = dA_{\rm H}$, and $f_{\rm V}$,
$f_{\rm H}$ are the holomorphic gauge-coupling functions.
In the following we will review the mechanism suggested in ref.~\cite{Witten:1985bz,Verlinde:2007qk} to 
obtain an effective Lagrangian \eqref{Lagr_split}
for a linear combination $A$ of $A_{\rm H}$ and $A_{\rm V}$.
Namely, as we will check for our scenario later on, the dimensional reduction to 
four space-time dimensions can induce a coupling term of the form 
\beq
\label{StueckelbergMass_U(1)}
\cL(\cC)= \cC\wedge \dd (q\, A_{\rm V}+e\, A_{\rm H}) +\frac{1}{2  \mu^{2}}  |\dd \cC|^2\ , 
\eeq 
where $\cC$ is a massless two-form field and $e,q$ are $U(1)$ charges. 
In four space-time dimensions the massless two-form $\cC$ is dual to 
a scalar $\rho$. The effective Lagrangian obtained by 
dualizing \eqref{StueckelbergMass_U(1)} is given by 
\beq \label{HiggsMass}
\cL(\rho) =\tfrac{1}{2} \mu^2 |d\rho +q\, A_{\rm V}+e\, A_{\rm H}|^2\ .
\eeq
As a consequence of the standard Higgs mechanism for $\rho$ 
there is a heavy and a massless mass eigenstate 
\beq
\label{Light_U(1)}
  A^{\rm h}=q\,A_{\rm V}+e\,A_{\rm H}\ ,\qquad \qquad q\, A= q\, A_{\rm V}-e\, A_{\rm H} \ .
\eeq 
Here we normalized the light $U(1)$ such that the visible sector fields have the
same charge under $2 A$ and $A_{\rm V}$. 
To make the Higgsing more explicit, we note that in an $\cN=1$ supersymmetric theory the
scalar $\rho$ in \eqref{HiggsMass} will combine with a second real scalar $v$ to
form the bosonic content of a chiral multiplet. This chiral multiplet is then 
absorbed by the $U(1)$ vector multiplet $A^{\rm h}$ to form a massive vector
multiplet $(A^{\rm h}, \xi(v))$, where $\xi(v)$ is the dynamical
Fayet-Iliopoulos term. The scalar $\rho$ has been absorbed by the
gauge transformation $ A^{\rm h} \rightarrow A^{\rm h} - d\rho$ in \eqref{HiggsMass}.
In a string compactification the heavy mass state $A^{\rm h}$ has typically a mass of order string 
scale and can be integrated out. The resulting effective Lagrangian takes the 
desired form \eqref{Lagr_split}. 
Up to corrections suppressed by the mass of $A^{\rm h}$, the effective
gauge-coupling of the massless $U(1)$ is given by 
\beq \label{effective_f}
   4 \R f = \R f_{\rm V} +(q/e)^2\ \R f_{\rm H} \ .
\eeq
The effective gaugino mass $\tilde M$ for the light $U(1)$ is computed 
by inserting \eqref{effective_f} into \eqref{Bare_Bino_mass}.
Note that the above results are only true at leading order, and further
corrections will arise due to integrating
out the massive $U(1)$ vector multiplet. The effects on the pattern of 
soft supersymmetry breaking terms can be evaluated rather
general as shown in ref.~\cite{ChoiScrucca}, following earlier works \cite{heavyU(1)}.

\subsubsection{Outline of the scenario \label{OutlineScenario}}

To be more explicit we now consider type IIB string theory compactified on
a six-dimensional  manifold $\cM_6$, which we choose to be a non-K\"ahler 
deformation of a Calabi-Yau 3-fold, such that the four-dimensional 
effective theory is still an $\cN=2$ supergravity theory.
The supersymmetry will be further reduced to $\cN=1$ by an  
orientifold projection and the inclusion of space-time filling D-branes \cite{review_flux}.
For many orientifold projections the spectrum of this theory contains 
a number of $U(1)$~vector multiplets with 
vectors arising from the R-R four-form $C_4$. For simplicity, 
let us concentrate on one such vector field $A_{\rm H}$ 
and its magnetic dual $\tilde A_{\rm H}$. Both arise as Kaluza-Klein
modes of $C_4$ by integrating 
\beq \label{def_A_H}
  A_{\rm H} = \int_\cA C_4\ ,\qquad \qquad \tilde A_{\rm H} = \int_\cB C_4\ ,
\eeq
where $\cA, \cB$ are three-dimensional submanifolds in $\cM_6$ with 
$\cA\cap \cB=1$. That $C_4$ contains both $A_{\rm H}$ and $\tilde A_{\rm H}$ 
is due to the fact that its field strength $F_5=\dd_{10} C_4$ needs to obey the ten-dimensional 
self-duality constraint $F_5 = * F_5$.
For appropriate $\cA,\cB$ the self-duality of $F_5$ implies 
the electro-magnetic duality between $A_{\rm H}$ and $\tilde A_{\rm H}$.
As indicated by the notation, the vector $A_{\rm H}$ will correspond to the hidden sector 
$U(1)_{\rm H}$ of section~\ref{rev_U(1)}.

The visible MSSM-like sector is realized on stacks of space-time filling 
D3 or D7 branes \cite{review_Dbrane,Douglas:1996sw,singMSSM,Model_F}. Later
on, we will mostly focus on intersecting branes on a del Pezzo four-cycle $S$. 
The resulting four-dimensional effective gauge theory generically contain a number of
$U(1)$ factors which arise, for example, from the splitting of the $U(N)$ gauge groups
on a stack of $N$ branes into $U(N) = SU(N) \times U(1)$. An appropriate combination 
of such $U(1)$ factors will provide a vector field $A_{\rm V}$, the visible 
sector $U(1)_{\rm V}$. In the MSSM $U(1)_{\rm V}$ has to coincide with the 
non-anomalous hypercharge when modeling the scenario of section \ref{HyperAnomaly}. 
In many intersecting brane models also additional $U(1)$ symmetries are induced and 
can be identified with the $U(1)'$ of the model in section \ref{U(1)'model}.

So far we introduced two decoupled sectors containing $A_{\rm H}$
and $A_{\rm V}$ respectively. In order to obtain an effective action 
of the form \eqref{Lagr_split}, with a common light $U(1)$ vector field 
 $A = qA_{\rm V}-eA_{\rm H}$, our string compactification 
should admit the couplings in the St\"uckelberg Lagrangian
\eqref{StueckelbergMass_U(1)} or equivalently  \eqref{HiggsMass}
to a four-dimensional two-form $\cC$ or its dual scalar $\rho$. In our orientifold compactification,
both $\cC$ and $\rho$ are obtained as Kaluza-Klein modes of the R-R four-form $C_4$ as 
\beq \label{def-rhoC}
   \cC = \int_{\Sigma} C_4\ .  \qquad \qquad  \rho = \int_{\tilde \Sigma} C_4\ . 
\eeq
Here $\Sigma$ and $\tilde \Sigma$ are two- and four-dimensional submanifolds 
of $\cM_6$ respectively, fulfilling $\Sigma \cap \tilde \Sigma=1$. The 
self-duality of $F_5$ implies the four-dimensional duality of $\rho$ and 
$\cC$.

In the next step we have to specify the topology of 
$\cM_6$ and the properties of $\cA,\cB$ in \eqref{def_A_H} as well 
as $\Sigma,\tilde \Sigma$ in \eqref{def-rhoC} in order that 
$A_{\rm V}+A_{\rm H}$ is massive in 
our string compactification. 
Let us first discuss 
the gauge term of the form \eqref{HiggsMass} induced for $A_{\rm H}$. 
A more detailed analysis of this gauging can be found in section \ref{Compact}.
In Calabi-Yau reductions, where $\cA,\cB$ and $\Sigma,\tilde \Sigma$ are harmonic 
cycles,  $\rho$ and $\cC$ do not couple to  
the vector field $A_{\rm H}$. However, we can couple $\rho$ to $A_{\rm H}$ 
if we impose the topological conditions 
\beq \label{boundaryCond}
   \partial \tilde \Sigma = e\, \cA\ , \qquad \qquad \partial \cB = e\, \Sigma\ ,
\eeq
i.e.~for some constant $e$ the cycles $e\cA$ and $e\Sigma$ should be boundaries of 
$\tilde \Sigma$ and $\cB$ respectively. The kinetic terms for $\rho$
arise from the ten-dimensional term $\int F_5 \wedge * F_5$. The 
first topological relation in \eqref{boundaryCond} yields that 
\beq \label{cov_H}
  \int_{\tilde \Sigma} F_5 =  d\rho + \int_{\partial \tilde \Sigma} C_4  = d\rho + e\, A_{\rm H}\ .
\eeq
Here we split $\dd_{10}=\dd_{4}+\dd_{6}$, applied Stokes Theorem in $6$
dimensions and used \eqref{def_A_H} and \eqref{def-rhoC}. In other words, the
condition \eqref{boundaryCond} 
implies that the scalar $\rho$ gets gauged by the $U(1)_{\rm H}$ exactly as 
in \eqref{HiggsMass}. Note that \eqref{boundaryCond} together with the 
fact that $\rho$ sits in the same supermultiplet as a K\"ahler modulus 
implies that $\cM_6$ cannot be a K\"ahler manifold as we show 
in section \ref{Compact}.

\begin{figure}[!ht]
\begin{picture}(200,120)
\put(80,10){\includegraphics[height=3.8cm]{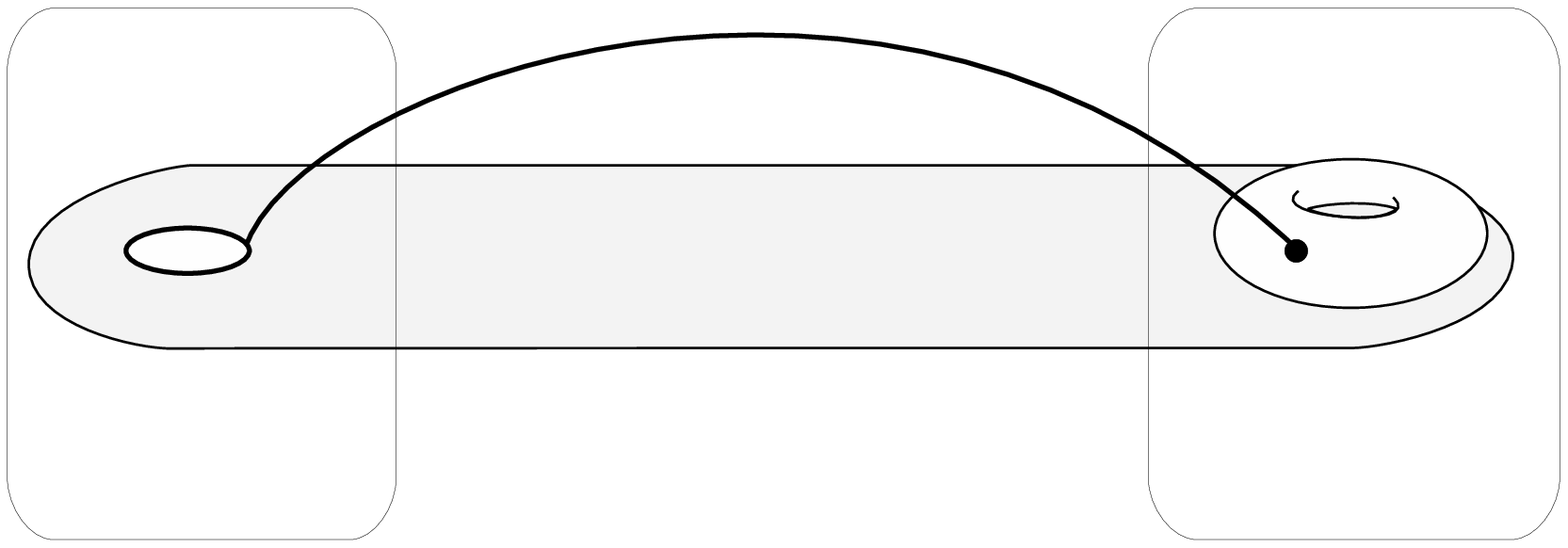}
}
\put(130,55){$\cA$}
\put(215,90){$\cB$}
\put(215,55){$\tilde \Sigma$}
\put(325,60){$\Sigma$}
\put(345,82){$S$}
\put(95,30){\small local flux}
\put(95,17){\small geometry}
\put(180,17){\small orientifold bulk}
\put(300,30){\small visible sector}
\put(300,17){\small on D-branes}
\end{picture}
\vspace*{-.5cm} 
\caption{\small  \label{Regions} Non-Calabi-Yau space with two local geometries. The three-dimensional chain $\cB$
  reaching through the orientifold bulk has a two-dimensional boundary
  $\Sigma=\partial \cB$ in the visible MSSM sector on a four-cycle $S$. The four-dimensional 
  chain $\tilde \Sigma$ reaching through the bulk has a three-dimensional 
  boundary $\cA = \partial \tilde \Sigma$ in the hidden flux geometry. }
\end{figure}

In a next step we also need to couple the vector field
$A_{\rm V}$ to the R-R fields $\rho$ or $\cC$. Here it is natural
to concentrate on the coupling to $\cC$ via the Chern-Simons 
action of the space-time filling D7 branes. As we will discuss in section \ref{D7_sector}, 
this action contains a term of the form 
\beq \label{D7coupling}
   \int_{{\rm \cW}_{7,1}} C_4 \wedge \cF \wedge \cF = q \int_{\mathbb{M}_{3,1}} \cC \wedge dA_{\rm V} + \ldots\ ,
\eeq
where $\cF$ is the field strength on the D7 brane world-volume ${\cW}_{7,1}$ 
and $\mathbb{M}_{3,1}$ is our space-time. Here $q$ is an 
induced  D5 charge arising from  
fluxes on the D7 brane. The coupling to $\cC$ defined in \eqref{def-rhoC} 
can be non-vanishing if $\Sigma$ is in the world-volume of the D7
brane.\footnote{Strictly speaking, it will be sufficient if there exists a
two-cycle in the world-volume of the D7 brane which can support D5 charge
which induces the coupling \eqref{D7coupling}. As for the compact orientifold 
example presented in section \ref{eF1} this can be the case even though $\Sigma$ itself
is not a curve in the D7 world-volume.}
The coupling \eqref{D7coupling} is a St\"uckelberg mass 
term of the form \eqref{StueckelbergMass_U(1)}. Since $\rho$ and $\cC$ are dual in four 
dimensions we can thus combine \eqref{cov_H} and \eqref{D7coupling} showing
that $\rho$ is gauged as in \eqref{HiggsMass}.
Precisely as in \eqref{Light_U(1)} this determines a massless and a heavy
linear combination $A,A^{\rm h}$.
A schematic overview of our set-up is presented in figure \ref{Regions}.

To complete our string set-up we now have to discuss how the fermion $\lambda$
in the vector multiplet $(A,\lambda)$ receives a bare mass from an hidden sector F-term
as in \eqref{Bare_Bino_mass}. Here the key point is, that the gauge coupling function $f_{\rm H}$ of 
$A_{\rm H}$ is holomorphic in the complex structure deformations of $\cM_6$  \cite{GL1}. 
The complex structure deformations appear in 
the flux superpotential $W_{\rm flux}$ induced by 
R-R and NS-NS three-form fluxes $F_3$ and $H_3$.
Denoting  by $\tau$ the complex dilaton axion and 
by $\Omega$ the holomorphic three-form on $\cM_6$
we have 
\beq \label{GVWsuperpot}
   W = \int_{\cM_6} G_3 \wedge \Omega\ , \qquad \qquad G_3 = F_3 -\tau H_3\ .
\eeq
The superpotential depends holomorphically on the complex structure deformations 
through $\Omega$. The corresponding scalar potential 
can admit minima with non-vanishing 
F-terms, which thus induce a bare mass for $\lambda$.

%
%

\section{$U(1)$ Mediation in Non-K\"ahler Compactifications \label{Compact}}

In this section we study a detailed realization of $U(1)$
mediated supersymmetry breaking in a Type IIB orientifold
compactification. We discuss the construction 
of the internal non-K\"ahler geometries in section~\ref{nonKaehler_res}. 
The orientifold projection as well as the four-dimensional $\cN=1$ effective action
and its characteristic functions are studied in section~\ref{nonKaehler_ori}.
The visible sector on a del Pezzo surface is introduced in section~\ref{D7_sector}, 
while the hidden flux sector breaking supersymmetry 
is discussed in section~\ref{hiddenSusy}. The main focus of this section 
is the study of the four-dimensional effective action in terms of the topological 
data and relations of the internal manifold. Most of the explicit geometric constructions are 
postponed to section~\ref{geometric_real}.

\subsection{Non-K\"ahler resolutions \label{nonKaehler_res}}

Let us focus on the compactification of type IIB string 
theory on a six-dimensional manifold $\cM_6$. In order 
to realize a $U(1)$ coupling between the hidden flux and 
the visible brane sector as in section \ref{U(1)_string}, 
$\cM_6$ cannot be a Calabi-Yau manifold. However, in order 
that the four-dimensional effective theory still possesses some
supersymmetry, $\cM_6$ needs to admit a globally defined three-form $\Omega$ 
and a two-form $J$, which define an $SU(3)$ structure on $\cM_6$ such that 
\beq \label{JOmega_prop}
  J \wedge \Omega =0 \ ,\qquad \quad J \wedge J \wedge J = c\,\Omega \wedge \bar
  \Omega \neq 0\ ,
\eeq
for some complex constant $c$ on $\cM_6$ \cite{CS,GLMW,Lopes Cardoso:2002hd}. 
$\Omega$ and $J$ are the analogs of the holomorphic three-from 
and K\"ahler form on a Calabi-Yau manifold. However, on a general $SU(3)$
structure manifold both $dJ$ and $d\Omega$ can be non-vanishing.

In our set-up we wish to deviate as little as possible 
from the Calabi-Yau geometry in order to keep 
in good approximation the powerful calculational 
tools of the $\cN=2$ special geometry. Therefore, we will restrict 
ourselves to manifolds which are still \textit{complex}, but can be \textit{non-K\"ahler}.
In terms of $J,\Omega$ this implies \cite{CS}
\beq \label{non-Kaehler_cond}
   d \Omega =0 \ ,\qquad \qquad dJ = \cW_3 \ , \qquad \cW_3 \wedge J = 0\ ,
\eeq
where $\cW_3$ is the three-form parameterizing the obstruction 
of $\cM_6$ being K\"ahler. The condition \eqref{non-Kaehler_cond} 
will be realized by a non-K\"ahler resolution of a singular Calabi-Yau
manifold~\cite{Werner,Candelas:1990rm,Chuang:2005qd}.

Let us consider conifold transitions between a Calabi-Yau manifold $Y$ 
to a manifold $\cM_6$.
In such a topological transition one or more cycles $\cA_i$ of $S^3$ topology are shrunken to a
node and resolved by exceptional two-cycles $\Sigma_a$ with $S^2$ topology.
There are global restrictions which need to be satisfied in order that $\cM_6$
remains K\"ahler. 
A well-known example of K\"ahler transitions are transitions between Calabi-Yau 
manifolds $Y \rightarrow Y'=\cM_6$. These occur if the  shrinking three-cycles obey a
number of relations in homology. For example, let us consider 
a transition in which $k$ $S^3$'s, denoted by $\cA_i$, shrink to nodes 
which are subsequently blown up to  $k$ $S^2$'s. 
Suppose  $\delta$ is the number of homological relations 
\begin{equation}
\sum_{i=1}^k\, c^{i}_j\ {\cal A}_i
=\partial \tilde \Sigma_j\ , \qquad j=1,\ldots,\delta \ ,
\label{homrelations}
\end{equation}  
with constant coefficients $c^{i}_j$ and four-chains $\tilde \Sigma_j$.
Since the independent $A$-cycles correspond locally one to one 
to variations of the complex structure, 
one has to fix $k-\delta$ complex structure moduli to
create the $k$ nodes in $Y$. This implies $h^{2,1}(Y)-h^{2,1}(Y')=k-\delta$. 
Further, in order for $Y'$ to be K\"ahler with $\dd J=0$ there must  be 
$k-\delta$ homology relations among the $k$ exceptional $S^2$'s. If this is the  
case a Calabi-Yau transition from $Y$ to $Y'$ exists and $h^{1,1}(Y')-h^{1,1}(Y)=\delta$.  
As we will argue next, one can also violate the K\"ahler condition in a simple 
and controlled way and thus construct non-K\"ahler manifolds.

Let us start with the simplest example of a transition to a non-K\"ahler manifold.
From the above discussion, we infer that one can never 
shrink a single $S^3$ cycle $\cA$, which is non-trivial  
in homology, i.e.~$k=1$, $\delta=0$ in \eqref{homrelations}, and resolve the 
singular geometry by an $S^2$ cycle $\Sigma$ 
such that the resulting geometry is K\"ahler.
The reason is that the non-trivial three-cycle $\cB$, which is symplectic dual 
to $\cA$ with $\cA \cap \cB=1$, develops a puncture at the nodal singularity as 
$\cA$ shrinks to zero size. It is easily seen from the local geometry near the intersection 
of $\cA,\cB$ that the exceptional two sphere $\Sigma$ becomes then a boundary of $\cB$. 
As a consequence as soon as one resolves the node to $\Sigma$
with finite size $v = vol( S^2)$ 
one gets \cite{Werner,Candelas:1990rm}
\begin{equation} \label{nonkaehler}
0\neq v=\int_{\Sigma} J=\int_{\partial \cB} J=\int_{\cB} \dd J\ .
\end{equation}  
The non-vanishing $dJ$ implies that the manifold
$\cM_6$ cannot be  K\"ahler.

\begin{figure}[!ht]
\begin{picture}(100,50)
\put(130,0){\includegraphics[height=1.2cm]{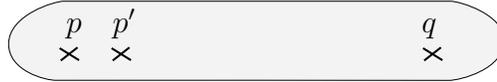}
}
\put(154,20){$p \quad p' \hspace{3.8cm} q$}
\end{picture}
\vspace*{-.5cm} 
\caption{\small  \label{Nodes} Divisor $\tilde \Sigma$ with three nodes
  $p,p',q$.}
\end{figure}

These transitions to non-K\"ahler manifolds can be generalized to yield the set-ups 
suggested in section~\ref{U(1)_string}. 
A convenient way to achieve this is to both resolve and deform nodes located on a
divisor $\tilde \Sigma$. Let us illustrate such a process on a simple
example which can be generalized easily to more complicated situations. 
We assume that we have a singular Calabi-Yau manifold with 
three nodes $p,p',q$ located on some divisor $\tilde \Sigma$ as depicted in figure~\ref{Nodes}.
After deforming all singularities into three-spheres $\cA,\cA',\tilde \cA$ they obey 
\beq \label{Cohom_cond}
 \cA + \cA' + \tilde \cA= \partial \tilde \Sigma\ ,
\eeq
which is \eqref{homrelations} for $\delta = 1$ and $k=3$. 
If \eqref{Cohom_cond} is the only condition in homology relating the $\cA$'s
there will exist two symplectic dual non-trivial three-cycles $\cB,\cB'$ with $\cA\cap \cB=\cA'\cap \cB'=1$
and $\cA\cap \cB'=\cA'\cap \cB=0$ such that 
\beq \label{intersect}
   \cB \cap (\cA+\cA'+\tilde \cA)= \cB' \cap (\cA+\cA'+\tilde \cA)=0\ .
\eeq

\begin{figure}[!ht]
\begin{picture}(100,80)
\put(120,0){\includegraphics[height=2.8cm]{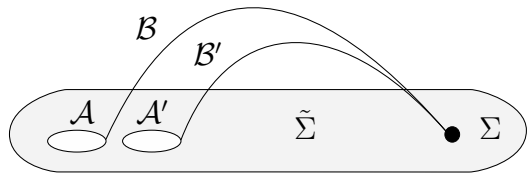}
}
\put(170,60){$\cB$}
\put(192,50){$\cB'$}
\put(145,28){$\cA \quad\ \cA'$}
\put(230,22){$\tilde \Sigma$}
\put(300,22){$\Sigma$} 
\end{picture}
\vspace*{-.5cm} 
\caption{\small  \label{23Spheres} Divisor $\tilde \Sigma$ with boundary
  $\cA+\cA'$ and the three-chains $\cB,\cB'$ with
  boundary $\Sigma$.}
\end{figure}

Clearly, we could also resolve any number of  nodes into two-spheres. This corresponds
to a geometric transitions of some or all of the three-cycles in \eqref{Cohom_cond}.
As discussed above, the blow-up process can preserve $dJ=0$ if there are homological relations 
between the shrinking three-cycles, since the K\"ahler volumes of the resolving $S^2$'s can
cancel. However, we can also decide 
to only resolve the third node $q$ into a two-sphere $\Sigma$, while
deforming the nodes $p,p'$ into three-spheres $\cA,\cA'$. This is shown 
in figure \ref{23Spheres} and corresponds
to a non-K\"ahler transition, with a homological relation that becomes
very useful for our geometric engineering of supersymmetry breaking. 

The evaluation of $dJ$ is now similar to
\eqref{nonkaehler}. The condition \eqref{intersect} implies 
that the two $\cB,\cB'$ as well as $\tilde \Sigma$ have a boundary 
\beq \label{boundary_rel}
  \partial \cB=\partial \cB' = \Sigma\ ,\qquad \qquad \partial \tilde \Sigma =
  \cA + \cA'\ .
\eeq
Performing the integral of $dJ$ over $\cB,\cB'$ we thus find 
\beq \label{def-v}
   \tfrac12 \int_{\cB+\cB'} d J =\int_{\Sigma} J = v \ ,\qquad  \qquad  \int_{\cB-\cB'} d J =0\ . 
\eeq 
This implies that $dJ$ is non-vanishing and the resulting manifold $\cM_6$ is
non-K\"ahler. However, we can ensure that $d\Omega = 0$ on $\cM_6$
by canceling the two holomorphic volumes of the deformed $S^3$'s. 
More precisely, due to the second relation in \eqref{boundary_rel} we need to obey the condition
\beq \label{dOmega=0}
  0=\int_{\tilde \Sigma} d\Omega =\int_{\cA + \cA'} \Omega = X^1 + X^2\ ,
  \qquad \qquad X^1= \int_{\cA} \Omega\ ,\quad X^2= \int_{\cA'} \Omega\ ,
\eeq 
such that $X^1 = -X^2$. As we will see in section \ref{nonKaehler_ori}, this condition 
can be consistently imposed together with an orientifold involution of $\cM_6$
which maps $\cA$ to $\cA'$.
 
Let us also comment on the construction of figure \ref{23Spheres}
from the point of view of the two local geometries. 
In order to do that we can imagine that we zoom into either of the two regions of figure
\ref{23Spheres}. One region contains a patch around 
the small three-spheres $\cA,\cA'$, while the second region
is the patch containing the two-sphere $\Sigma$. Effectively, in this process 
we obtain two non-compact geometries by scaling the connecting chains 
$\cB,\cB'$ and $\tilde \Sigma$ to be infinitely large. In the
local geometries we are still able to identify the compact two- and
three-cycles $\Sigma$ and $\cA,\cA'$, while the information about the chains is
lost in performing the local limit. In particular, this implies that the
condition \eqref{def-v}, \eqref{dOmega=0} cannot be evaluated in the local
geometry and require global information about $\cM_6$. In non-compact
geometries one has to \textit{choose} the dual non-compact cycles $\tilde \Sigma,\cB,\cB'$ 
by fixing appropriate boundary conditions. When patched into a
compact space one has to demand that they fulfill the conditions \eqref{boundary_rel}.

\subsection{The effective action of non-K\"ahler orientifolds \label{nonKaehler_ori}}

Compactifying type IIB string theory on the non-K\"ahler manifold $\cM_6$
leads to a four-dimensional effective $\cN=2$ supergravity theory \cite{GLMW}.
An appropriate orientifold projection will reduce this further to $\cN=1$~\cite{BG,Koerber:2007xk}. 
For set-ups with O3/O7 planes the orientifold 
projection $\cO = \Omega_p (-1)^{F_L} \sigma$ contains 
an involution $\sigma$ obeying 
\beq \label{transJOmega}
  \sigma^* J  = J \ , \qquad \qquad \sigma^* \Omega = - \Omega\ ,
\eeq
where $J$ and $\Omega$ are the globally defined two and three-form obeying
\eqref{JOmega_prop} and \eqref{non-Kaehler_cond}.
The orientifold symmetry $\sigma$ also splits the cohomologies 
into positive and negative eigenspaces $H^{n} = H^{n}_+ \oplus H^{n}_-$
with dimensions $b^{n}_\pm = \dim H^{n}_\pm$. 
For simplicity, we will restrict to manifolds $\cM_6$ with 
$b^{2}_- = 0$.
To remain in the orientifold theory the ten-dimensional NS-NS
B-field $B_2$ and the R-R forms $C_0,\, C_2,\, C_4$ 
have to transform under the involution $\sigma$ as
\beq \label{trans_BC}
   \simga^* B_2 = -B_2\ ,\qquad \quad \sigma^* C_p = (-1)^{p/2}\, C_p\ .
\eeq

In performing the Kaluza-Klein reduction to four space-time dimensions, 
we have to expand all ten-dimensional fields in forms on $\cM_6$ transforming 
with the appropriate sign under $\sigma^*$. 
Note that in a consistent compactification on a non-K\"ahler space this   
will also involve non-harmonic representatives \cite{GLMW,BG,Koerber:2007xk,Robbins:2007yv}. In the following 
we generalize the discussion of section \ref{nonKaehler_res} to the case of $N$
non-K\"ahler resolutions. We denote the small resolving two-spheres by
$\Sigma_i$, while the deforming pairs of three-spheres are denoted by
$\cA_i,\cA_i'$, with $i=1,\ldots,N$. 
Later on, we will realize all $\Sigma_i$ in a four-cycle $S \in H_{4}(\cM_6)$ 
on which the visible gauge theory is modeled. In other words we will 
identify $\Sigma_i \in H_{2}(S)$, while there will be homological relations 
among the $\Sigma^i$ within $\cM_6$. The compact three-cycles $\cA_i,\cA_i'$ 
will support the hidden flux geometry. They are non-trivial in the local Calabi-Yau geometry
around the hidden singularity. 
The chains connecting the two- and three-cycles in $\cM_6$ are
denoted by $\tilde \Sigma^i$ and $\cB^i,\cB'^i$. In order to connect the hidden and 
visible sector we demand that they obey 
\beq \label{multi_chain}
  \partial (\cB^i+\cB'^i) = e^{ij} \Sigma_j\ ,\qquad \partial \tilde \Sigma^i
  = e^{ij}(\cA_j + \cA'_j)\ ,
\eeq 
for some constant matrix $e^{ij}$ of rank $N$. The equation
\eqref{multi_chain} is the generalization of \eqref{boundary_rel}. 
In summary, we associate to each non-K\"ahler resolution 
\beq \label{submani}
   (\Sigma_i,\tilde \Sigma^i)\ ,\qquad (\cA_i,\cA'_i,\cB^i,\cB'^i)\ ,\qquad
   i=1,\ldots,N\ .
\eeq
The $\cN=1$ orientifold involution $\sigma^*$ introduced in
\eqref{transJOmega} is chosen such that 
\beq \label{action_oncycles}
   \sigma^* \Sigma_i = \Sigma_i\ ,\qquad \quad \sigma^* \tilde \Sigma^i = \tilde
   \Sigma^i\ , \qquad \quad \sigma^* \cA_i = \cA_i'\ ,\qquad \quad \sigma^* \cB^i = \cB'^i\ ,
\eeq
Note that the first two conditions are not necessarily true point-wise
for all points on $\Sigma_i,\tilde \Sigma^i$, such that 
$\Sigma_i,\tilde \Sigma^i$ are  
not necessarily entirely inside an orientifold plane. Finally, there can be cycles 
$\Gamma_a \in H_2(S)$ in the visible sector region transforming as 
\beq \label{negative_forms}
   \sigma^* \Gamma_a = -\Gamma_a\ .
\eeq
Since we demand that $b^{2}_- = 0$ the cycles $\Gamma_a$ will be trivial in $\cM_6$. 
Note that a convenient way to invariantly characterize the introduced basis is provided by using 
relative homology.\footnote{See e.g.~\cite{Karoubi:1987} for an introduction to relative homology.}
For example, the relative homology group $H_3(\cM_6/S)$, will by definition contain the elements
of $H_3(\cM_6)$ as well as three-chains with boundaries on~$S$.
Using the chains introduced in \eqref{submani} and \eqref{negative_forms} we will be able to 
determine the spectrum of the four-dimensional effective 
theory. It is then shown that the Kaluza-Klein modes associated to the
non-harmonic chains, summarized in Table~\ref{tab:Chains}, 
appear as massive scalar fields in the effective theory.

\begin{table}[htbp]
  \centering
  $
  \begin{array}{|c|c|c|c|}
     \hline
      \rule[-0.2cm]{0cm}{0.6cm}  \text{chain}& \text{dim}_{\bbR}&
      \text{relations}& \sigma\text{-parity}\\
     \hline
     \rule[-0.25cm]{0cm}{0.8cm}\hspace*{.2cm} (\Sigma_i,\cB^i+\cB'^i)
      \hspace*{.2cm}&\hspace*{.2cm} (2,3)\hspace*{.2cm} & 
      \hspace*{.2cm} \partial (\cB+\cB')=e \Sigma \hspace*{.2cm}&+ \\
      \hline
     \rule[-0.25cm]{0cm}{0.8cm} (\tilde \Sigma^i,\cA_i + \cA'_i) & (4,3) & \partial \tilde \Sigma
  = e (\cA+\cA') &+ \\
\hline
   \rule[-0.25cm]{0cm}{0.8cm} \Gamma_a & 2 & \text{boundary} & - \\
\hline
  \end{array}
  $
  \caption{\label{tab:Chains}{Non-harmonic chains used in the Kaluza-Klein reduction.}}
\end{table}

\subsubsection{The complex structure sector}

Let us first discuss the four-dimensional fields associated to the 
holomorphic three-form $\Omega$. Recall that  we demanded 
in \eqref{non-Kaehler_cond} that our manifold $\cM_6$ is still complex. 
This implies that the space of complex structure
deformations will admit a similar structure as in the Calabi-Yau case. 
We introduce the periods $(X^K,\cF_K)$ as
\beq \label{exp_Omega}
    X^K = \int_{\cA^K_-} \Omega\ ,\qquad  \quad \cF_{K} = \int_{\cB_K^-} \Omega\ , \qquad \quad K=0,\ldots , b^{3}_- -1\ ,
\eeq
where $(\cA^K_-,\cB_K^-)$ is a real symplectic basis of $H_3^-(\cM_6)$.
$\Omega$ and its periods depend holomorphically on $b^{3}_- -1$ complex
 structure deformations $z^k$. 
By the local Torrelli Theorem  the complex structure 
deformations can be mapped locally one to one to the projective 
space spanned by the periods $X^K$.
In special coordinates this map is given by $z^k=X^k/X^0$. Note that 
all cycles in \eqref{exp_Omega} have to be in the negative eigenspace of $\sigma^*$
due to \eqref{transJOmega}. In particular, if $\Omega$ is integrated over a positive 
cycle $\cA_i+\cA_i'$ one has
 \beq   \label{special_coords}
    X_+^i(z) \equiv X^0 z^i_+ = \int_{\cA_i+\cA'_i} \Omega = 0\ .
 \eeq
 This condition is in accord with \eqref{dOmega=0} where it was imposed to ensure 
 that $d\Omega=0$. Similar constrains have to be imposed for integrals over elements in 
 $H_3^+(\cM_6)$. Note however, that there are $U(1)$ vectors arising from the integrals 
 of the R-R four-form $C_4$ over these positive cycles. In accord with \eqref{trans_BC} 
 one has
 \beq \label{def-AH}
    A_{\rm H}^i =  \int_{\cA_i+\cA'_i} C_4 \ , \qquad \qquad A^{\kappa} = \int_{\cA_\kappa^+} C_4\ ,  \qquad \quad \kappa=1,\ldots , b^{3}_+\ ,
 \eeq
 where $\cA_\kappa^+$ is a basis of $H_3^+(\cM_6)$. Note that the vectors arising from integrals over the 
 symplectic dual cycles of $\cA_i+\cA'_i$ and $\cA_\kappa^+$ do not contain new degrees of freedom due to 
 the self-duality of $C_4$ as discussed in section \ref{U(1)_string}. While in the underlying $\cN=2$ theory 
 the complex structure deformations pair with the $U(1)$ vectors into $\cN=2$ vector multiplets the orientifold 
 splits these into $\cN=1$ chiral and vector multiplets. One thus finds $b^3_-$ chiral multiplets with 
 bosonic scalars $z^k$ and $b^3_+ + N$ vector multiplets with bosonic components $A_{\rm H}^i, A^{\kappa}$.

The $\cN=1$ characteristic functions for $z^k$ and the vectors $A_{\rm H}^i, A^{\kappa}$
are readily determined following \cite{GL1,BG}. The K\"ahler potential for
 the complex structure deformations $z^k$ takes the well-known form 
\beq \label{def-Kcs}
  K_{\rm cs}(z,\bar z) = - \log \big[-i\int \Omega \wedge \bar \Omega \big]\ .
\eeq
The gauge-kinetic coupling functions of the $U(1)$ vectors $A_{\rm H}^i, A^{\kappa}$
will be denoted by $f^{\rm H}_{ij},f_{\kappa \lambda}$ and $f_{\kappa i}$. They are holomorphic functions of the complex
structure deformations $z^k$. In general there will be a kinetic mixing between the $A_{\rm H}^i$ and $A^{\kappa}$ 
through $f_{\kappa i}$. For simplicity, we will assume that this mixing is small such that the massless 
$A^{\kappa}$ decouple from the $A_{\rm H}^i$.\footnote{Note that the presence of kinetic 
mixing can have an interesting effect on the visible phenomenology as discussed recently in a
string theory context in ref.~\cite{Abel:2008ai} (see also the references therein).} 
Our main focus will be on the vectors $A_{\rm H}^i$ with gauge-coupling function given by
\beq \label{f_H(z)}
   f^{\rm H}_{ij}(z)  = -i  \frac{\partial^2 \cF}{\partial {z_+^i}
     \partial {z_+^j}} \Big|_{z^i_+ = 0}\ ,
\eeq
where $\cF(z)$ is the $\cN=2$ pre-potential depending on all complex structure 
deformations $z_+^i,z^k$ defined after \eqref{exp_Omega} and in
\eqref{special_coords}. As explained in more detail in ref.~\cite{GL1}, 
the $\cN=1$ gauge-kinetic coupling function is obtained by first taking derivatives 
with respect to $z^i_+$ and then restricting to the orientifold locus $z^i_+=0$. We will 
argue in section \ref{hiddenSusy} that $K_{\rm cs}$ and $f^{\rm H}_{ij}(z) $ should 
be calculable at least near the local hidden singularity.

\subsubsection{The non-K\"ahler sector}

Let us now turn to the $\cN=1$ chiral multiples arising from $J$ and $B_2$. 
It was shown in ref.~\cite{GL1,BG} that the orientifold theory enforces a particular complex 
structure on the $\cN=1$ chiral field space which combines the NS-NS fields 
with the R-R fields into complex scalars. More explicitly, one introduces
\bea \label{def-tauGT}
  \tau &=& C_0 + i e^{-\phi} \ , \qquad \quad G^a = \int_{\Gamma_a} (C_2 - \tau B_2)\ , \nn \\
  T_{M} &=& -\int_{\tilde \Sigma^M}  e^{-B_2}  \wedge C^{\rm RR}  + i e^{-\phi}  \int_{\tilde \Sigma^M}\tfrac{1}{2} \big( J \wedge J -B_2 \wedge B_2\big)\ ,
\eea 
where $C^{\rm RR}=C_0 + C_2 +C_4$. The two-cycles $\Gamma_a$ are introduced in \eqref{negative_forms}, while the 
four-chains $\tilde \Sigma^M=(\tilde \Sigma^\alpha,\tilde \Sigma^i)$ consist of a basis $\tilde \Sigma^\alpha$ of $H^+_4(\cM_6)$ 
and the chains $\tilde \Sigma^i$ introduced in \eqref{submani}. 
Note that $\tau,G^a,T_M$ are nothing else then the integrals of $e^{-B_2}  \big(C^{\rm RR}  + i e^{-\phi} \,\R( e^{i J})\big)$ over 
one, two and four-chains respectively. These chains have to transform with definite signs under $\sigma^*$ to 
due to the transformation \eqref{transJOmega} and \eqref{trans_BC} of $J,B_2$ and $C_p$. 
The K\"ahler potential for $\tau,G^a,T_M$ is given by \cite{BG}\footnote{Strictly speaking this K\"ahler potential is 
valid only in the large volume limit of $\cM_6$ and will receive corrections once cycles in $\cM_6$ become small.}
\beq \label{K_pot_q}
 K_{\rm q}=  - 2 \log \big[e^{-2\phi} \cV \big] \ , \qquad \cV = \tfrac{1}{3!} \int_{\cM_6} J \wedge J \wedge J\ ,
\eeq
where $\cV$ is the string-frame volume of the compact manifold $\cM_6$. The total K\"ahler potential 
for the bulk modes is then given by $K = K_{\rm cs}+K_{\rm q}$, with $K_{\rm cs}$ as in \eqref{def-Kcs}.

In order to evaluate the K\"ahler metric, one needs to evaluate the K\"ahler potential \eqref{K_pot_q}
as a function of the $\cN=1$ coordinates $\tau,G^a$ and $T_M$ defined in \eqref{def-tauGT}.
This is in general very hard,  in particular since the internal manifold is not K\"ahler and we cannot apply 
$\cN=2$ special geometry. However, the derivatives of $K_{\rm q}$ can be evaluated using the work 
of Hitchin \cite{Hitchin} as done in \cite{BG}. Firstly, one notes that $K_{\rm q}$ only depends 
on the dilaton $e^{-\phi}$ and $J$. It was shown in \cite{Hitchin} that $e^{-2\phi} \cV$ is a well-defined 
functional of $e^{-\phi}\R(e^{iJ})$ as long as this form is closed. This is indeed the case, since we 
imposed $dJ \wedge J = 0$ in \eqref{non-Kaehler_cond} and we are in the orientifold limit where $\phi$ is constant 
on $\cM_6$. This ensures that $K_{\rm q}$ can be evaluated as function of $\I\, \tau$ and $e^{-\phi}  \int_{\tilde \Sigma^M} J \wedge J $.
In order to translate the latter into a dependence on $\I\, T_M$ one needs to compensate the $B_2$ in $\I\, T_M$ using 
$\I\, G^a$. This can be done consistently on each divisor $\tilde \Sigma^M$ and in particular for the four-cycle $S$
providing the visible sector. More generally, it was shown in ref.~\cite{Hitchin} that $e^{-2\phi} \cV$ can be evaluated 
as a function of $e^{-\phi}\R(e^{-B_2+iJ})$ as long as this form is closed under $d+H_3\wedge$, where 
$H_3 =\langle dB_2 \rangle$ is the NS-NS three-form flux.

For the evaluation of the K\"ahler metric it is essential that the K\"ahler potential does 
not depend on the R-R forms $C_0,C_2$ and $C_4$. In particular, note that in
contrast to $J\wedge J$ the four-form $C_4$ is not closed. This non-closedness results in a gauging of the scalars $T_i$
since 
\beq \label{gauging_T}
    \int_{\tilde \Sigma^i} d(e^{-B_2} \wedge C^{\rm RR}) = d_4 \R T_i + \int_{\tilde \Sigma^i} d_6 C_4 = d_4 \R T_i + e_{ij} A^i_{\rm H}\ ,
\eeq
where $d_4$ and $d_6$ are the differentials in the visible and compact dimensions respectively, and we have 
used \eqref{multi_chain} in evaluating the last equality. This implies that 
we have to replace the ordinary derivatives in
the four-dimensional kinetic terms for $T_M=(T_\alpha,T_i)$ by the covariant derivatives 
\beq  \label{cov_T}
 \cD T_\alpha = dT_\alpha\ ,  \qquad \quad  \cD T_i = dT_i + i e_{ij} A^j_{\rm H}\ ,
\eeq
where $A^i_{\rm H}$ are the $U(1)$ vector fields in \eqref{def-AH}. The gauging 
of $T_i$ will induce a D-term providing a potential for the modes arising from 
non-harmonic forms. Note that the $\cN=2$ analog of the gauging \eqref{gauging_T} has been 
studied in refs.~\cite{GLMW}.\footnote{In this case, the analysis of the kinetic terms is more complicated 
since the R-R fields reside in the quaternionic geometry and the application 
of Hitchins work is likely to be more involved.}

\subsubsection{The scalar potential}

So far we discussed the kinetic terms for the vectors and scalars in 
the $\cN=1$ effective action. Even in the absence of background fluxes, we
expect that a potential is generated for the compactification on $\cM_6$.
In particular it should give a mass to the fields arising from the
non-K\"ahler deformation resolving the singular Calabi-Yau space. 
Such a potential will arise precisely from the gauging \eqref{cov_T} of the 
scalars $T_i$. Recall that the $\cN=1$ 
scalar potential is of the form 
\beq \label{general_N=1}
  V = e^K \big(G^{I \bar J} D_I W \overline{D_{J}W} -3 |W|^2 \big) +
  \tfrac{1}{2} (\R f)^{KL}\, D_K D_L\ ,
\eeq 
where $(\R f)^{KL}$ is the inverse of the real part of the gauge-coupling
function $f_{KL}$.
Focusing on \eqref{cov_T} we note that $e_{ij}$ is invertible and 
all $N$ complex fields $T_i$ are gauged. This implies 
that there are $N$ D-terms $D_{i}^{\rm H}$ induced in the  potential
\eqref{general_N=1}. In the case at hand we evaluate using \eqref{def-tauGT}, 
\eqref{K_pot_q} and \eqref{multi_chain} that \footnote{%
Recall the general formula for the D-term of a $U(1)$ symmetry,
$K_{I \bar J} \bar X^{\bar J}_k =  \partial_{I} D_k$, where $X^J$ is the Killing vector of the $U(1)$ symmetry given 
by $\delta M^I = \Lambda^k X_k^J \partial_J M^I$. }
\beq \label{D_H}
   D^{\rm H}_i = -ie_{ij} \, \partial_{T_i} K = 4  e^{K_{\rm q}/2}\, e^{-\phi}
   \int_{\cB^i + \cB'^i}dJ\ ,
\eeq
where $K_{\rm q}$ is given in \eqref{K_pot_q}. The simple form of this D-term
arises due to the fact that there are no other scalars in the spectrum charged 
under $A^i_{\rm H}$. One might have suspected that, at least at the special
locus where some of the three-cycles become very small, additional states
charged under $A^i_{\rm H}$ arise. These would correspond to light D3-branes
wrapped on the vanishing cycles and contribute light hypermultiplets in the
underlying $\cN=2$ theory \cite{Strominger:1995cz}. However, such states are actually absent
if there is a R-R flux on the shrinking three-cycle \cite{Intriligator:2003xs}.

As expected for a non-K\"ahler reduction, the D-term \eqref{D_H} and the gauging 
\eqref{cov_T} will induce a mass for the complex scalars $T_i$. However, 
it remains to discuss the scalar potential for the fields $G^a$ defined in \eqref{def-tauGT}
as integrals of $C_2$ and $B_2$ over the negative cycles $\Gamma_a$. 
Such a mass term will arise from the the reduction of the ten-dimensional term 
\beq \label{mass_termG}
  \tfrac14 \int_{10} e^{\phi}\,  G_3 \wedge *_{10} \bar G_3 = \int_4 *_4 \mathbf{1}\ m_{ab}\, G^a \bar G^b + \ldots\ ,
\eeq 
where $G_3 = F_3 -\tau H_3$ contains the  field strengths of 
$C_2$ and $B_2$. The mass $m_{ab}$ will depend on the size of 
the three-chains with boundary cycles $\Gamma_a$. The 
mass term \eqref{mass_termG} can also be translated into an
$\cN=1$ potential  \eqref{general_N=1} and will arise from a 
superpotential. This $W$ is of the form \eqref{GVWsuperpot}, i.e.~given by
\beq \label{ext_GVW}
    W = \int_{\cM_6} G_3 \wedge \Omega\ ,
\eeq 
where $G_3$ is the internal part of the field strengths of $C_2,B_2$. 
This $W$ should contain terms linear in $G^a$ arising form the exact forms $d_6 (C_2-\tau B_2)$, 
where $d_6$ is the differential on $\cM_6$.
However, in order to derive \eqref{mass_termG} from \eqref{ext_GVW} one has 
to allow variations of $\Omega$ which are non-closed and hence leave the 
class of complex manifolds. The consideration of this extended class of spaces
is crucial since $\cM_6$ is compact. On the local non-compact Calabi-Yau
geometries of $\cM_6$ one cannot move $d_6$ onto $\Omega$ in \eqref{ext_GVW}. Therefore, the 
non-compact case allows to model the non-complex geometries by changing the
boundary conditions for the non-compact cycles (see the recent discussion in~\cite{Aganagic:2008qa,Hollands:2008cs}). 

For the discussion of $U(1)$ mediation the fields $G^a$ will be of no importance. 
We will make use of the fact that they are massive and can be fixed to vevs where $\cM_6$ is complex. 
The flux superpotential \eqref{ext_GVW}  can also fix the dilaton $\tau$ as well as the complex structure moduli. We will 
assume that the only light complex structure moduli arise from the hidden singularity discussed 
in section \ref{hiddenSusy} and trigger supersymmetry breaking. 
Finally, in order to stabilize all moduli of the theory, one should also fix the 
moduli $T_\alpha$ supersymmetrically by, for example,
using the mechanisms proposed in \cite{KKLT}.

\subsection{The visible sector on D-branes \label{D7_sector}}

In this section we discuss the inclusion of space-time 
filling D-branes which provide the 
visible sector in the effective Lagrangian \eqref{Lagr_split}.
We apply and generalize the results of refs.~\cite{JL,Haack:2006cy,Buican:2006sn}
to the orientifold compactifications of section \ref{nonKaehler_ori}. 
The four-dimensional theory will admit 
charged chiral matter fields if this sector consists of a number of
intersecting D-branes. In order to provide the 
ground for the examples considered in section \ref{geometric_real}, 
we will concentrate on branes at singularities of the internal space.
The local geometry allows the D-branes to split up into 
several intersecting fractional branes which can be engineered to 
yield a semi-realistic visible 
spectrum \cite{review_Dbrane,Douglas:1996sw,singMSSM}.\footnote{For our 
purposes, it will not be crucial to model a fully realistic MSSM.}

The singularities we will consider are obtained by shrinking 
a four-cycle in $\cM_6$ to a point. 
In the following we will exemplify the general strategy 
for the case of del Pezzo surfaces. These are either the surfaces $\cB_n$
which are obtained by blowing up $\bbP^2$ on $n$ generic points, or $\bbP^1 \times \bbP^1$. 
The del Pezzo $\cB_n$ has Hodge numbers 
$h^{(1,1)}=n+1$ and $h^{(2,2)}=h^{(0,0)}=1$, with all other numbers 
vanishing, so that the Euler number is 
\begin{equation}
\chi(\cB_n)=\int_{\cB_n} c_2=3+n \ .
\label{eulerdelpezzo}
\end{equation} 
Using Hirzebruch-Riemann-Roch \cite{BPV} gives for the arithmetic 
genus of a surface $S$:  $\chi_0=\sum_{p=0}^{{\rm dim}(S)}(-1)^p h^{p,0}
=\frac{1}{12}\int_S (c_1^2+c_2)$, and since $\chi_0(\cB_n)=h^{0,0}=1$ we get
\begin{equation}
K^2=\int_{\cB_n} c_1^2=9-n \ .
\end{equation} 
One calls $K^2$ also the degree of the del Pezzo surface.
A base of homologically nontrivial two-cycles in $\cB_n$ consists of the class 
of lines $l$ in $\bbP^2$ as well as the $n$ exceptional curves with classes 
$e_i$ corresponding to the blow-ups. The intersection numbers are 
$l^2=1$ and $e_i^2=-1$ with all other intersections vanishing.

As an alternative basis on can use the degree zero sublattice of $H_2(\cB_n,\bbZ)$
which has zero intersection with the canonical class 
\begin{equation} 
K = -3l + \sum_i e_i\ .
\label{canonicalclassdelpezzo}
\end{equation} 
For $n\ge 3$ this lattice is identified with the root lattice of the groups 
$\cE_n$, where for $n=6,7,8$ $\cE_n=E_n$ are the exceptional groups $E_n$,
$\cE_5=D_5$, $\cE_4=A_4$ and  $\cE_3=A_2\oplus A_1$. 
Defining the simple roots by 
\begin{equation}  \label{delpezzolattice}
\alpha_i=e_i-e_{i+1}\ , \quad i=1,\ldots, n-1\ ,\qquad \qquad \alpha_n=l-e_1-e_2-e_3 
\end{equation} 
it is immediate that the intersection matrix of $\alpha_i$ is 
given by the negative Cartan matrix $-C^{ij}$ of the corresponding Lie
algebra. The intersection matrix $\cK^{IJ}$ of the cycles 
$\Sigma_I=(K,\alpha_i),\ I=0,\ldots,n$ is thus given by
\beq \label{int_matrix_delP}
   \cK^{00}=9-n\ ,\qquad \cK^{ij}= -C^{ij}\ ,\qquad \cK^{0i} = 0\ .
\eeq

Let $C$ be a curve in the del Pezzo surface. Then its degree $deg(C)$ and its 
arithmetic genus $g$ reads
\beq \label{deg_g}
  deg(C)=-K.C\ ,\qquad \qquad g=\tfrac{1}{2}(C.C+K.C)+1\ .
\eeq
Lines on the del Pezzo surface fall into representations of the Weyl-Group. 
To understand the geometry 
of the embedding of del Pezzo surfaces in Calabi-Yau spaces it 
is useful that the number of some lines will appear as rational 
instantons of degree one in the classes realized globally in the 
embedding Calabi-Yau. For convenience of the reader we reproduce Table 3
of Ref.~\cite{Demazure}.

\begin{table}[htbp]
  \centering
  $
  \begin{array}{|c|rrrrrrrrr|}
     \hline
      \rule[-0.2cm]{0cm}{0.6cm}  \text{class}&\cB_1& \cB_2&\cB_3&\cB_4&\cB_5&\cB_6&\cB_7&\cB_8 &\\
     \hline
     \rule[-0.1cm]{0cm}{0.55cm} (0;-1)&1&2&3&4&5&6&7&8 &\\
    \rule[-0.1cm]{0cm}{0.55cm}  (1;1^2)&&1&3&6&10&15&21&28&\\
     \rule[-0.1cm]{0cm}{0.55cm} (2;1^5)&&&&&1&6&21&56&\\
    \rule[-0.1cm]{0cm}{0.55cm}  (3;2,1^6)&&&&&&&7&56&\\
     \rule[-0.1cm]{0cm}{0.55cm} (4;2^3,1^5)&&&&&&&&56&\\
     \rule[-0.1cm]{0cm}{0.55cm} (5;2^6,1^2)&&&&&&&&28&\\
    \rule[-0.2cm]{0cm}{0.55cm}  (4;3,2^7)  &&&&&&&&8&\\
     \hline
     \rule[-0.1cm]{0cm}{0.55cm}  \text{Total no.} &1&3&6&10&16&27&56&240&\\
\hline
  \end{array}
  $
  \caption{\label{tab:(0,1,1)}{Number of lines on the $\cB_n$ del Pezzo
      surfaces. The coefficients $(a;b_1,\ldots b_n)$ describing the classes
      are given w.r.t.~the generators $(l;-e_1,\ldots,-e_n)$}.}
\end{table}

In the del Pezzo surface $\cB_n$ there are $n+1$ K\"ahler parameters associated with the 
volumes of the two-cycles $(K,\alpha_i)$. In general, for $\cB_n$ 
in some compact Calabi-Yau space $\iota:\, \cB_n \hookrightarrow Y$, not all of the two-cycles will descend to
K\"ahler moduli of $Y$. The number of associated K\"ahler moduli is determined 
by the rank of the map 
\beq \label{PiCY}
 \Pi:\  H_{2}(\cB_n) \ \rightarrow \ H_{2}(Y)\ , \qquad \quad
 \Pi(\Sigma,\omega) = \int_{\Sigma} \iota^* \omega\ , 
\eeq
where $\Sigma \in H_{2}(\cB_n)$ and $\omega \in H^2(Y)$. Using the orientifold 
involution $\sigma$ we can split $H_2(\cB_n)$ into eigenspaces
$H^+_{2}(\cB_n) \oplus H^-_{2}(\cB_n)$ with basis $(\Sigma_\kappa^+, \Sigma_a^-)$ and accordingly
decompose the intersections \eqref{int_matrix_delP} as
\beq  \label{split_intersection}
  \cK^{\kappa\lambda}_+ = \Sigma_\kappa^+ \cdot \Sigma_\lambda^+ \ , \qquad \cK^{ab}_- = \Sigma_a^-
  \cdot \Sigma_a^-\ ,
\eeq
with the mixed intersections $\Sigma_\kappa^+ \cdot \Sigma_a^-$ vanishing.
Since $\iota$ commutes with $\sigma$, one can split $\Pi$ into maps from the
$\sigma$-eigenspaces $H^\pm_{2}(\cB_n)$ to $H_{2}^\pm (Y)$.

Note that in the case that the internal manifold is a non-K\"ahler space $\cM_6$ 
as in section \ref{nonKaehler_ori} the map $\Pi$ will no longer identify cohomology
elements of $\cM_6$ and $\cB_n$, but should also include the boundary cycles
$\Sigma_i$ and $\Gamma_a$ of Table \ref{tab:Chains}.
Correspondingly, the new map can
be of higher rank. Following our considerations of section \ref{nonKaehler_ori} the cycles 
$\Sigma_i,\Gamma_a$ will precisely be in the four-cycle $S=\cB_n$ supporting 
the visible sector, i.e.
\beq
  \Sigma_i \in H^+_2(\cB_n)\ ,\qquad \quad \Gamma_a \in H^-_2(\cB_n)\ .
\eeq
In summary, there will be two types of cycles: two-cycles in
$H_2(\cB_n)$ which are non-trivial in $H_2(\cM_6)$, as well as 
two-cycles which are homologically trivial in $\cM_6$. Recall that we 
are taking $b^2_-=0$, such that all elements of $H_2^-(\cB_n)$ are trivial 
in $\cM_6$. In section \ref{delPezzogeometries} we will discuss how one can count the number 
of cycles of the various types, and also show that on a del Pezzo 
$H^2_-(\cB_n)$ is always non-empty if $\sigma$ acts non-trivially 
on $\cB_n$.

Let us now wrap branes on the surface $\cB_n$. The non-trivial zero-, two- and
four-cycles in $\cB_n$ can support fractional space-time
filling D-branes. We will denote the field strength of the $k^{\rm th}$ stack 
of fractional branes by $\cF^k$. The number of $D3$, $D5$ and $D7$ branes are
encoded by the charge vector 
\beq
  \text{ch}(\cF^k)\ \cong\ (r_k,p_{k}^I,q_k)\ .
\eeq
The $D7$ charge is the rank $r_k$ of $\cF^k$, the 
D5 charge is captured by fluxes $ p^I_k \cong( p_{k}^\kappa, p_{k}^a)$ on two-cycles $\Sigma_I \cong (\Sigma_\kappa^+ , \Sigma_a^-) $ in
$\cB_n$, and the D3 charge is encoded by the instanton number $q_k$.
Explicitly, these are given by 
\beq \label{def-p}
  p_{k}^\kappa = \int_{\Sigma^+_\kappa} \text{Tr} (\cF^k)\ ,
\qquad\quad p_{k}^a = \int_{\Sigma^-_a} \text{Tr} (\cF^k)\ ,
\qquad\quad  q_k = \tfrac12 \int_{\cB_n} \text{Tr} (\cF^k \wedge \cF^k)\ .
\eeq
This form of the charge vector can be inferred from the 
coupling of the $k^{\rm th}$ brane stack to the R-R forms $C^{\text{RR}}$ via the 
Chern-Simons action\footnote{Note that in general there will be corrections to $S^{CS}$
given by $\sqrt{\hat A_T/\hat A_N}$, where $\hat A_T,\hat A_N$ are the $\hat A$ genera of the tangent and normal bundle
to the brane world-volume. These will equivalently appear in the definition \eqref{def-tauGT} of the 
$\cN=1$ coordinates.}
\beq \label{CS-action}
   S^{\rm CS} = \int_{\mathbb{M}_{3,1} \times \cB_n} \iota^*
   C^{\text{RR}}\wedge e^{\cF^k - \iota^* B_2}\ .
\eeq
where $C^{\text{RR}} = C_0 + C_2 +C_4$ is the sum of the R-R potentials as in 
\eqref{def-tauGT}. It turns out to be convenient to 
also  define the topological numbers $ q_{k\, \kappa} = p_k^\lambda\, \cK_{\lambda \kappa}^+$
and $q_{k\, a} = p_k^b\, \cK_{b a}^-$,
where we lower the indices with the inverses of  
the intersection matrices given in \eqref{split_intersection}.

Expanding the Chern-Simons action and integrating over the four-cycle 
$\cB_n$ one encounters the term of the 
form $\I f_k^{\rm V}\,  F^k
\wedge F^k$, which contains the four-dimensional part $F^k$ of the field
strength $\cF^k$. This term determines the imaginary part of the 
gauge-coupling function $f_k^{\rm V}$. Using the fact that $f_k^{\rm V}$ is holomorphic 
in the $\cN=1$ coordinates defined in section \ref{nonKaehler_ori}, it is determined to be 
\beq \label{f_V}
  f^{\rm V}_k =-i(T_{\cB_n} -  q_{k\, a}  G^a - q_k\, \tau)\ ,
\eeq 
where $T_{\cB_n},G^a,\tau$ are the $\cN=1$ coordinates defined in \eqref{def-tauGT}.
In particular, $T_{\cB_n}$ is the K\"ahler structure coordinate \eqref{def-tauGT} corresponding to the 
del Pezzo four-cycle~$\cB_n$ and takes the form
\beq
  T_{\cB_n} = -\rho_{\cB_n} + i e^{-\phi} \vol_{\cB_n}  - \frac{1}{2 (\tau - \bar \tau)} \cK^-_{ab} G^a (G-\bar G)^b\ ,
\eeq
where $\rho_{\cB_n}$ is the R-R axion and $\vol_{\cB_n}$ is the volume of $\cB_n$.

In addition to the gauge-coupling function, there are also
further couplings of the R-R four-form to the 
gauge-theory.
The expansion of the Chern-Simons action \eqref{CS-action} yields the term
\beq  \label{dF5-correction}
       \int_{\cB_n}  \text{Tr} (\cF^k \wedge  \cF^k) \wedge  \iota^* C_4 = 
      q_{k\, M}\  F^k \wedge
       \cC^M  +  \ldots \ ,
\eeq
where $\cC^M=(\cC^i,\cC^\alpha)$ are the four-dimensional two-forms 
dual to the R-R four-form scalars in $\R\, T_M$ defined in \eqref{def-tauGT}.
Clearly, $q_{k\, A}$ vanishes if $\tilde \Sigma^M$ in \eqref{def-tauGT} has 
no intersection with the two-cycle in $\cB_n$ supporting the $D5$ charge.
The contribution \eqref{dF5-correction} to the effective action is of the
form \eqref{StueckelbergMass_U(1)} and hence induces a gauging 
of the scalars $\R\, T_M$. This implies
that the covariant derivatives \eqref{cov_T} are modified to 
\bea \label{fullDT}
   D T_\alpha &=& dT_\alpha + i q_{k\, \alpha} A^k_{\rm V}\ , \\
   D T_{i}  &=&  dT_i + i (e_{ij}\, A_{\rm H}^j + q_{k\,i}\, A^k_{\rm V} )\ ,\nn 
\eea
where $A^k_{\rm V}$ is the visible $U(1)_k$ factor on the $k^{\rm th}$ D-brane stack.
Note that the fields $T_i$ can thus be gauged by both the hidden and visible
sector gauge-fields. As in eqn.~\eqref{Light_U(1)}, we can now
identify heavy and light mass eigenstates proportional to 
$e_{ij}\, A_{\rm H}^j + q_{a\,i}\, A^a_{\rm V}$
$e_{ij}\, A_{\rm H}^j - q_{a\,i}\, A^a_{\rm V}$.
As explained in section \ref{rev_U(1)}, the linear combinations appearing in
\eqref{fullDT} can become heavy via the Higgs mechanism and can be integrated out. 
The remaining light vector fields then couple to both the hidden flux geometry as well
as the visible gauge theory on the fractional branes and can mediate
supersymmetry breaking.

We are now in the position to derive the D-term potential 
for the whole configuration. In addition to the bulk D-term 
\eqref{D_H} we have additional contributions
\beq \label{Dvisible}
   D^{\rm V}_a = 4 e^{K_{\rm q}} \int_{\cB_n} \text{Tr}(\cF_a) \wedge \iota^* J -
   \sum_i Q_{i}^{(a)} |\phi_i|^2\ ,
\eeq
where $Q^{(a)}_i$ are the $U(1)_a$ charges of the canonically normalized
matter fields $\phi_i$.
The full D-term potential is now obtained by inserting 
\eqref{f_H(z)}, \eqref{D_H} as well as \eqref{f_V} and \eqref{Dvisible} into \eqref{general_N=1}.

\subsection{The hidden flux geometry and supersymmetry breaking \label{hiddenSusy}}

In this section we study the hidden sector supporting a supersymmetry 
breaking flux background. 
Supersymmetry breaking by background fluxes has been investigated 
since the advent of flux compactifications \cite{review_flux}.  It was shown that 
warped Calabi-Yau compactifications with flux superpotential \eqref{GVWsuperpot} 
admit supersymmetric vacua at points in the moduli space where the complex flux $G_3$ is a $(2,1)$ form. 
If this condition is violated supersymmetry appears to be broken spontaneously.
Unfortunately, in the full compactification such a conclusion can be too quick. 
The effects of the supersymmetry breaking fluxes have to be sufficiently small and
localized to ensure persisting control over the effective four-dimensional theory 
and moduli stabilization. Moreover, supersymmetry breaking fluxes can
backreact on the internal geometry and render the 
compact manifold to be no longer Calabi-Yau.

A simple way to obtain meta-stable non-supersymmetric flux vacua in
\textit{local} Calabi-Yau geometries has been studied in ref.~\cite{ABSV}. 
We will use orientifolds of the set-ups \cite{ABSV} to model
a supersymmetry breaking sector. In order to do that, we zoom into local 
regions $\hat X, \hat X'$ of the compactification manifold $\cM_6$.
Let us assume that $\hat X$ and $\hat X'$ are locally Calabi-Yau 
and get exchanged under the geometric orientifold involution $\sigma$. 
In order to not worry about cross couplings connecting fields on $\hat X$ 
and its orientifold image $\hat X'$ we will demand that these regions are away from the orientifold 
planes in $\cM_6$. This will allow us to work on $\hat X$ keeping in mind that 
there exists an identical copy $\hat X'$.

Considering type II string theory on the non-compact Calabi-Yau $\hat X$ forces us 
to decouple gravity. Nevertheless, we can study the moduli space of 
complex structure deformations $\underline{S}=(S^i),\ i=1,\ldots,h^{2,1}(\hat X)$ by analyzing the variations 
of the holomorphic three-form $\Omega(S)$.
Let us focus on the cases where $\hat X$  contains
a number of compact three-cycles $\cA_i$ with $S^3$ topology. 
A simple example $\hat X$ is obtain as deformations of a fibered $A_1$ 
singularity \cite{Cachazo:2001jy}. Such a local geometry is given by a complex 
equation in $\bbC^4$ of the form 
\begin{equation} 
u^2+ w^2+v^2+p_m(t)^2+ f_{m-1}(t|\underline{S})=0\ , \qquad p_m = g \prod_i (t
-a_i)\ ,
\label{localsing1_A1}
\end{equation}
where the subscripts indicate the degree of the polynomial functions in 
$t$.  The local geometry (\ref{localsing1_A1}) can be described as follows. 
If $f_{m-1}(t|\underline{S})=0$ one obtains $m$ nodal singularities of
the local form (\ref{node}) at $u=w=v=0$ and the roots $a_i$, $i=1,\ldots,m$
of $p_m(t)$. 
The $f_{m-1}(t|\underline{S})$ destroys the 
factorization in $t-a_i$ and deforms the nodes into $m$ $S^3$'s denoted by $\cA_i$. These $S^3$'s are 
homologically distinct and their size can be parameterized by $m$ independent complex structure 
deformations~$\underline{S}$.
In order to make contact to the discussion of section \ref{nonKaehler_ori} 
we note that $\cA_i$ should include the cycles $\cA_i$ 
introduced in \eqref{submani}. However, in full analogy to the discussion of the visible 
sector in section \ref{D7_sector}, the relations \eqref{multi_chain} only arise through the embedding of 
$\hat X$ and $\hat X'$ into the global non-K\"ahler space $\cM_6$.

The fact that $\hat X$ and its orientifold image are Calabi-Yau allows us to 
use $\cN=2$ special geometry to describe the moduli space spanned by $\underline S$.
One introduces the non-compact cycles $\cB^i$ which are the symplectic 
duals to the compact three-cycles $\cA_i$ in $\hat X$. This is possible 
if one introduces a cutoff $\Lambda_0$ to regulate integrals over $\hat X$.
The periods of $\Omega$ are thus given by
\beq
  S^i   = \int_{\cA_i} \Omega\ , \qquad \quad \partial_i \cF = \int^{\Lambda_0}_{\cB^i} \Omega\ ,
\eeq
where $\partial_i \cF= \partial_{S^i}\cF$, and special geometry ensures the
existence of a holomorphic pre-potential $\cF(\underline{S})$. 
For the fibered $A_1$ singularity \eqref{localsing1_A1} the B-periods $\partial_{i}\cF$ 
have been computed in ref.~\cite{Cachazo:2001jy}. At leading order they take the simple 
form 
\beq
  2 \pi i \partial_{i} \cF = S^i \big[\log \Big(\frac{S^i}{p_m'(a_i)
    \Lambda^2_0} \Big) -1 \big] + \sum_{i\neq j} S^j \log
  \Big(\frac{\Delta^2_{ij}}{\Lambda^2_0} \Big) +\ldots\ .
\eeq
Here $p_m'(a_i)$ is the first derivative of $p_m$ introduced in \eqref{localsing1_A1}
evaluated at the root $a_i$, and $\Delta_{ij}=a_i -a_j$ is the complex
distance between the nodes in the singular $\hat X$.

In the 
$\cN=2$ theory the complex scalars $S^i$ sit together with $U(1)$ vectors $A^i=\int_{\cA_i} C_4$ in vector multiplets. 
Note that on the orientifold image $\hat X'$ we can similarly introduce the periods 
$(S'^i,\partial'_i \cF)$ on the image cycles $(\cA_i',\cB'^i)$. 
The orientifold projection \eqref{transJOmega}, \eqref{trans_BC} 
ensures that the $\cN=2$ vector multiplets $(S^i,A^i)$ and $(S'^i,A'^i)$
split into $\cN=1$ chiral multiplets with complex scalars $s^i=(S^i-S'^i)/2$
and $\cN=1$ vector multiplets with $A^i_{\rm H}=A^i+A'^i$ ($A^i-A'^i=0$).
As in \eqref{special_coords} the orientifold locus is given by $S^i_+ =(S^i+S'^i)/2= 0$, while 
the condition $\partial \cF/\partial S^i_+ = 0$  automatically arises on the orientifold locus
due to the symmetries of $\cF$. 
The orientifolded four-dimensional effective theory will 
be a rigid $\cN=1$ supersymmetric theory.

Restricting the theory to the orientifold moduli space parameterized by $\underline{s}$,
the $\cN=1$ metric remains rigid special K\"ahler with K\"ahler potential 
\beq \label{simple_Kpot}
   K(\underline{s},\underline{\bar s}) = \tfrac{i}{2} (s^j \partial_{\bar s^j} \bar \cF - \bar s^j \partial_{s^j}\cF)_{\underline{S}_+=0}\ .
\eeq
The K\"ahler metric is simply given by $K_{i\bj}= \I( \partial^2
\cF/\partial s^i \partial s^j)$ restricted to $\underline{S}_+ = 0$.
The holomorphic function $f^{\rm H}_{ij}(\underline{s})=-i \partial^2
\cF/(\partial S_+^i \partial S_+^j)$ for $\underline{S}_+=0$ is the gauge-kinetic coupling function 
of the $U(1)$ vectors $A^i_{\rm H}$ as in \eqref{f_H(z)}. 
In order to allow for $U(1)$ mediation the embedding of $\hat
X$ and $\hat X'$ into the compact space $\cM_6$ has to ensure that 
$A^i_{\rm H}$ combines with a visible $U(1)$ vector $A^i_{\rm V}$ into a light 
and massive eigenstate $A^i,A^i_{\rm h}$ as in \eqref{Light_U(1)}. Upon
integrating out $A^K_{\rm h}$ as in section \ref{generalU(1)}, only $A^K$
remains in the low energy theory and has an effective gauge-coupling $f_{KL}$ 
given at lowest order the sum of the visible and hidden $f$'s as in \eqref{effective_f}.

Let us turn to the scalar potential for $\underline{s}$.
In rigid $\cN=1$ supersymmetry it takes the
form 
\beq \label{V_special}
 V= K^{i\bj} \partial_{i} W \partial_{\bj} \bar W  + \tfrac12 (\R
 f)^{-1\, ij} D_i D_j \ ,
\eeq
where $\partial_i W= \partial_{s^i}W$, and $D_i$ is the D-term for the light $U(1)$ vector $A^i$.
The superpotential $W$ arises due to a non-trivial flux background and 
using \eqref{GVWsuperpot} is given by 
\beq \label{superspot_simple}
   W =2(\alpha_i\, s^i +N^i\ \partial_{s^i}\cF)_{\underline{S}_+=0} \ ,
\eeq
where the factor $2$ arises due to the fact that there is an orientifold image
of $\hat X$ in $\hat X'$.
The flux quanta appearing in  \eqref{superspot_simple}
are given by $N^i = \int_{\cA_i} F_3$ and $\alpha^i = -\int_{\cB^i} (F_3 - \tau H_3)$,
where $F_3$ and $H_3$ are R-R and NS-NS three-form fluxes.
Here $\tau$ is complex dilaton-axion \eqref{def-tauGT} which, in the compact set-up, can be
stabilized supersymmetrically by other
background fluxes \cite{GKP}.
As in section \ref{generalU(1)} the leading contribution to the 
gaugino masses $ \tilde M^{i j}$ of the light $U(1)$ vector multiplets is given by
the generalization of \eqref{Bare_Bino_mass} with \eqref{effective_f}. 
Therefore, if there exists a non-supersymmetric minimum of the potential
\eqref{V_special} the F-term $F^m$ will be non-vanishing and contribute to $\tilde M^{ij}$.

A few comments concerning the presented outset are in order. 
Firstly, note that the described flux background was argued to be large-$N$ dual to a
set-up where the $S^3$'s are replaced by $\bbP^1$'s via geometric transition. 
The fluxes $N_i$ correspond to the rank of the gauge-group of $N_i$ $D5$ 
or anti-D5 branes wrapped on the $i$th $\bbP^1$. The gauge-coupling function of the 
branes on the large-$N$ dual of $\hat X$ at the scale $\Lambda_0$ is given by $\alpha(\Lambda_0)=\alpha_i$ for all stacks 
of branes. Secondly, note that the orientifold projection maps D5 to anti-D5
branes with the identification $N^i = - N'^i$ and $\alpha_i = -\alpha_i'$,
since the fluxes $F_3$ and $H_3$ are odd under the orientifold involution.
This implies that if the gauge-theory obtained from $\hat X$ arises
from $D5$ branes only, the gauge-theory from $\hat X'$ is due to anti-D5 branes. 
The five-branes on $\hat X'$ have to wrap flopped $\bbP^1$'s similar to the 
recent discussion in ref.~\cite{Aganagic:2008qa}. 

It was shown in refs.~\cite{ABSV}, that on geometries \eqref{localsing1_A1}
one can indeed find meta-stable supersymmetry breaking flux vacua of
\eqref{V_special}. In order to simplify the discussion we will consider the 
cases $m=1$ in the fibered $A_1$ singularity \eqref{localsing1_A1}, such
that each of the local Calabi-Yau spaces $\hat X$ and $\hat X'$
only admits one $S^3$'s respectively.
We will determine the minima of the scalar potential
\eqref{V_special} in the situation of small D-terms. 
In order to do that we  have to specify the sign of the fluxes $N$ and $\I
\alpha$. If both have the same sign, there is a supersymmetric
vacuum at $\langle s \rangle = g \Lambda_0^2\,  e^{-2\pi i \alpha/N}$.
Clearly, the gaugino mass $\tilde M$ vanishes in this vacuum. The situation
changes as soon as one has opposite signs of $N$ and $\I \alpha$. In the case 
$N<0$ and $\I \alpha>0$ one finds a non-supersymmetric minimum of $V$ at
\beq
  \langle s \rangle = g \Lambda_0^2\,  e^{2\pi i \bar \alpha/|N|}\ ,  \qquad
  \partial_s W = \alpha-\bar \alpha\ .
\eeq
In this non-supersymmetric vacuum the gaugino mass $\tilde M$ is non-zero and evaluated
to be $\tilde M \propto \tilde g^2 |N|/ \langle s \rangle$, where $\tilde g$
is the gauge-coupling of the mediating $U(1)$.
Note that even though this computation is very explicit and can be performed
including higher corrections to the pre-potential it typically does not lead
to the right scales for $\tilde M$ and supersymmetry breaking. This can be traced back 
to the fact that in the compact settings the fluxes are actually quantized in units of $\alpha'$.
Since the orientifold already specifies an $\cN=1$ supersymmetry inside the 
underlying $\cN=2$ theory, a small breaking cannot occur as in \cite{ABSV}.

One expects that the scales can be made phenomenologically 
viable by placing the hidden singularity in a warped throat \cite{Klebanov:2000hb,GKP,Giddings,DST,Shiu:2008ry}.  
This implies that the ten-dimensional metric background is of the form $ds^2= e^{2A} ds^2_{4} + e^{-2A} ds_{6}^2$, 
where $ds^2_6(y)$ is the line element on $\cM_6$, 
and the warp factor $e^{2A}(y)$ is depending on the internal coordinates. It is straightforward to verify that 
the warp factor  cancels for the four-dimensional gauge-coupling function $f^{\rm H}$. 
It is also believed that the warp factor does not induce leading corrections to the flux superpotential.
However, there will be corrections to the $\cN=1$ K\"ahler potential. So far only the leading corrections have
been analyzed in ref.~\cite{DST}, $\delta K \propto |s|^{2/3}$. These dominate for small $|s|$ such that the 
mass $\tilde M$ is of the form $\tilde M \propto \tilde g^2 \langle | s |^{1/3} \partial_s W \rangle$. $\tilde M$
can be small if one finds non-supersymmetric vacua for sufficiently small $|s|$.

Let us end with a brief comment on an alternative to the route taken here. 
An interesting possibility is to model fluxed supersymmetry breaking in the hidden sector by using the supergravity 
backgrounds perturbed by anti-D3 branes in a warped throat~\cite{DeWolfe:2008zy,Kachru:2002gs}. 
The anti-D3 brane will induce a supersymmetry breaking flux as required to generate the gaugino mass $\tilde M$.
It would be interesting to analyze such set-ups on the level of the effective action and to 
study the resulting pattern of soft supersymmetry breaking terms. If
supersymmetry breaking arises in a warped throat, partial sequestering can take
place and one expects a mixing of gravity and gauge-mediated contributions to
generate the visible soft masses.

%
%

\section{Geometric Realizations \label{geometric_real}}

In this section we discuss the geometric tools to construct internal manifolds with 
a hidden and visible sector. 
These sectors are realized near singularities of Calabi-Yau manifolds which
become non-K\"ahler by the mechanism discussed in section \ref{nonKaehler_res}. 
We start with a warm-up by recalling some basics on nodal singularities in section \ref{simple_node}.
For the hidden supersymmetry breaking sector we consider singularities with
a number of  deformed $S^3$'s in section \ref{singularitieswithsmallresolutions}.
For the visible sector we study the geometry of del Pezzo surfaces and realize
them in simple compact Calabi-Yau spaces in section \ref{delPezzogeometries}
and \ref{numericalchange}. The more involved compact examples admitting 
both the hidden as well as a visible del Pezzo singularity are discussed in
section \ref{eF1} and appendix \ref{torictransitions}.

\subsection{Simple nodal singularities \label{simple_node}}

As a warm-up for the more involved singularities needed for the 
visible and hidden sector, we will 
first recall some basic facts about 
complete intersections and briefly discuss the simple 
example of the deformed or resolved conifold in quintic Calabi-Yau. 

Most known Calabi-Yau spaces $Y$ are given in terms of generically smooth 
embeddings 
\beq \label{Toric}
   P_i(\underline {x}, {\underline z})=0\ ,\qquad i=1,\ldots, r\ ,
\eeq
into a toric ambient space $T_\Delta$. The case $r=1$ corresponds 
to the special case of a hypersurfaces, while $r>1$ defines complete  
intersections. In \eqref{Toric} the ${\underline x}$ are the coordinates of $T_\Delta$ and  
${\underline z}$ are complex deformation parameters of the polynomials. 
Generically smooth means that \eqref{Toric} and 
\begin{equation}  \label{sing} 
\dd P_1\wedge \ldots \wedge \dd P_r= 0\ ,
\end{equation} 
have no solutions for generic values of the deformation parameters 
${\underline z}$ and any value of ${\underline x}$. More precisely, 
the tangent space of complex structure deformations $H^1_{\bar \partial}(TY)$ 
is given in this situation generically by the space of infinitesimal 
deformations ${\rm def} (\underline P)$ modulo the infinitesimal automorphisms 
${\rm aut}({T_\Delta},{\underline P})$ of the ambient space, 
which are compatible with \eqref{Toric}. The infinitesimal complex structure 
deformations can be extended to a global moduli space $\cM^{\rm cs}$ without obstruction.%
\footnote{The
realization as complete intersection in particular 
toric ambient spaces can obstruct sometimes elements in $H^1_{\bar \partial }(TM)$. 
For simplicity we focus on complex structure deformations  in
$H^1_{\bar \partial }(TM)$ that can be realized by ${\rm def}(\underline P)/{\rm aut}
({T_\Delta},{\underline P})$.}
As in section \ref{nonKaehler_ori} and with slight abuse of notation,
we  call the complex structure deformations $\underline{z}$.   
Note that solutions to (\ref{sing}) will generically exist in subloci 
of complex codimension one (or higher) in ${\cal M}^{\rm cs}$. At a codimension 
one locus   
the Calabi-Yau manifold $Y$ can acquire a singularity for special values of $\underline x$. 
The most generic singularity is a node and it will be instructive to discuss some 
of the basic concepts for this simple case. 

For the simplest realization of these concepts consider the famous quintic 
surface in the toric ambient space $\mathbb{P}^4$. The quintic surface 
has $101$ complex structure deformations $\underline{z}$.\footnote{Generically a quintic 
constraint will have $126$ deformation parameters. The dimension 
of ${\rm Aut}(\mathbb{P}^4,P)=\mathbb{C}^*\times PGL(5,\mathbb{C})$ is 
$25$ rendering the number of complex parameters to $101$.} For 
simplicity consider the smooth one parameter family of quintics 
$P=\sum_{i=1}^5 x_i^5-5 z \prod_{i=1}^5 x_i=0$. Here $z\in \mathbb{C}$ 
is an unobstructed complex structure deformation.  Generically the 
constraints $\dd P=0$ and $P=0$ have no solution, but at $z=1$ and $x_i=1$, $i=1,\ldots,5$ 
they have a solution, the conifold point. The conifold divisor $z=1$ 
is codimension one in the one parameter family
of quintic surfaces. 
We find that the local singularity is a node by expanding the defining
equation of the Calabi-Yau manifold $P=0$ in an affine chart  $x_5=1$ for 
small $\mu=1-z$ near the singularity, i.e.~at $x_i=1+\tilde u_i$, 
$i=1,\ldots, 4$ with small $\tilde u_i$. After a linear change of 
variables from $\tilde u_i$ to 
$(u,v,w,t)$, we can bring the local 
equation to the normal form 
\begin{equation}
f=u^2+v^2+w^2+t^2-\mu(S)=0\ , 
\label{conifold}
\end{equation}
The node occurs at $\mu(S)=0$, where the real three sphere 
$S^3$ given by the real equation (\ref{conifold}) is contracted  
to zero size. Switching on $\mu(S)$ deforms the node into an $S^3$, which can 
be parameterized by a complex structure deformation $S$.
Locally, 
$S$ is given by the special coordinate
$S=\int_{S^3}\Omega(\mu)$. In the following we will use the letter $S$ to
denote the complex structure deformations in the local non-compact geometries
such as \eqref{conifold}.

By a further transformation the node singularity \eqref{conifold} at $\mu=0$ can be brought in the form
\begin{equation} 
\phi_1\phi_2-\phi_3\phi_4 =0\ ,
\label{node}
\end{equation}   
and admits two kinds of small resolutions by an $\mathbb{P}^1$. Introducing
two new projective complex coordinates $(x,y)$ the smooth blown up geometry can be either 
described by (\ref{node}) and the equations
\begin{equation}
\left(\begin{array}{cc} 
\phi_1& \phi_3\\ 
\phi_4& \phi_2 \end{array} \right) \left(\begin{array}{c} 
x\\ 
y \end{array}\right)=0 
\label{res} 
\qquad \text{or} \qquad
\left(\begin{array}{cc} 
\phi_1& \phi_4\\ 
\phi_3& \phi_2 \end{array} \right) \left(\begin{array}{c} 
x\\ 
y\end{array} \right)=0 \ .
\end{equation} 
At the point $P$ given by $\phi_1=\phi_2=\phi_3=\phi_4=0$ the coordinates  $(x,y)$ describe a
$\mathbb{P}^1$, while outside the singular point $P$  $(x,y)$ can be eliminated to 
recover the  geometry (\ref{node}). The holomorphic map 
\beq \label{res_map}
    \pi: \quad \hat X\  \rightarrow\ X\ ,
\eeq 
from the smooth resolution $\hat X$ to the nodal variety $X$ is called 
the resolution map. It identifies $\hat X$ with $X$ outside $P$. 
The resolution is called small, because the exceptional set $\pi^{-1}(P)=\mathbb{P}^1$
is of complex codimension two in the threefold. This implies that it 
does not affect the canonical class. The modification is local and
holomorphic and does not affect the complex structure. 
As we discussed in section \ref{nonKaehler_res} the resolution is generically 
non-K\"ahler. We have recalled in section \ref{nonKaehler_res}, that even if there 
are $\delta$ homology relations (\ref{homrelations}) among $k$ shrinking three-cycles, 
there will also be many non-K\"ahler resolutions 
among the $2^k$ resolutions defined by (\ref{res}) if $k\le \delta$ \cite{Werner}.

\subsection{The hidden singularity}
\label{singularitieswithsmallresolutions}

In order to model the supersymmetry breaking hidden sector we have to 
built more complicated singularities. This can be achieved 
by going to higher codimension in the moduli space of a Calabi-Yau 
family  with a large number of complex structure deformations. 
A classification of the local Calabi-Yau 3-fold singularities, which allow a resolution with 
trivial canonical bundle, has only been achieved in the 
hypersurface case \cite{Reid}.  

\subsubsection{Creating an A-D-E singularity}

The most relevant examples for the hidden 
sector for us are non-compact Calabi-Yau manifolds with singularities,
which admit small $S^2$ resolutions or $S^3$
deformations. They are of A-D-E type and given by the equation
\begin{equation}
u^2+ g(w,v)+ t^{m h} =0\ ,
\label{smallade3fold}
\end{equation}    
where $g(w,v)$ and the dual coxeter number $h$
of the associated Lie algebra $\mathfrak{g}$ are listed in Table \ref{tab-LocalSing}.
The requirement for the existence of a small resolution is $m\in 
\mathbb{N}$. Of course, for \eqref{smallade3fold} embedded into a compact Calabi-Yau manifold 
there will be an upper bound on the rank of the (gauge) group associated to $\mathfrak{g}$
that is realizable.
\begin{table}[h]
\begin{center}
\begin{tabular}{| c | c | c |} \hline
   \rule[-0.2cm]{0cm}{0.7cm} {\small Lie Algebra} $\mathfrak{g}$ &
   {\small Polynom} $g(w,v)$ & {\small Dual coxeter no.} 
$h$ 
   \\ \hline
   \rule[-0.1cm]{0cm}{0.6cm} $A_r$ &
   $w^2+v^{r+1}$ & $r+1$\\ 
   \rule[-0.1cm]{0cm}{0.6cm} $D_r$  &
   $w(v^2+w^{r-2})$ & 
   $2r-2$\\ 
\rule[-0.1cm]{0cm}{0.6cm} $E_6$  &
   $w^3+v^4$ & $12$\\ 
\rule[-0.1cm]{0cm}{0.6cm} $E_7$  &
   $w(w^2+ v^3)$ &
   $18$\\
\rule[-0.3cm]{0cm}{0.8cm} $E_8$  &
   $w^3+v^5$ & $30$\\ \hline
\end{tabular}
\caption{\small \label{tab-LocalSing}
\textit{A-D-E singularities.}}
\end{center}
\end{table}

In the following we will study orientifold involutions on
geometries with $A_r$ singularities for $r>1$.
These are obtained by fibering the $A_r$ two-fold in Table \ref{tab-LocalSing}
over a plane $\bbC[t]$ parameterized by $t$.
The local equation of the Calabi-Yau space is now of the form~\cite{Cachazo:2001sg}
\begin{equation}
P=uw-\prod_{i=1}^{r+1} (v-v^i(t)) + f(v,t|\underline{S})=0\ , \qquad
\sum_{i=1}^{r+1} v^i(t) =0\ ,
\label{localssing2}
\end{equation}
where $v^i(t)$ are polynomials in $t$ and $f(v,t|\underline{S})$ is a
polynomial in $v,t$. The singular geometry is obtained by setting $f(v,t|\underline{S})=0$
and admits nodes located at $u=w=0$ and $v=v^i(t)=v^j(t)$ for $i\neq j$.
If the highest power in $t$ in this singular Calabi-Yau space is of the form
\eqref{smallade3fold} one finds $m r(r+1)/2$ nodes.
The function $f(v,t|\underline{S})$ encodes the 
normalizable deformations of the singular Calabi-Yau space and is of the form \cite{Cachazo:2001sg}
\beq \label{general_deform}
   f(v,t) = f_{m-1}(t) v^{r-1} + f_{2m-1}(t)v^{r-2} +\ldots+f_{(r-1)m-1}(t) v
   + f_{rm-1}(t)\ .
\eeq
The coefficients in $f_i$ correspond to the $m r(r+1)/2$ 
complex structure deformations $\underline{S}$ which deform the nodes into $S^3$'s. It is easy to see that 
the general form \eqref{localssing2} reduces to \eqref{localsing1_A1} for the $A_{1}$ singularity. In this case 
one has $v^1(t)=-v^2(t)=p_m(t)$ and only the first and last term $f_{m-1}(t)$
remain in \eqref{general_deform}.

\subsubsection{Orientifolds of A-type singularities}

We are now in the position to specify a local orientifold action on the
$A_{r}$ geometries \eqref{localssing2} for $r=2k$.
We first write \eqref{localssing2} as 
\beq \label{ori_f}
   uw - (v-v^0(t)) \prod_{i=1}^k (v-v^i(t)) \cdot \prod_{j=1}^k (v-\tilde v^j(t)) + f(v,t|\underline{S})=
   0\ .
\eeq
Let us briefly discuss two holomorphic orientifold symmetries of
\eqref{ori_f}. The simplest case is a $\sigma_1$ which inverts 
one coordinate $t \mapsto -t$, while keeping all other directions 
invariant. This involution preserves an $O7$ plane at $t=0$, and we 
need to demand that it exchanges $v^i$ with $\tilde v^i$ and preserves $v^0,f$.
A second possibility $\sigma_2$ is to map $(t,v,u,w)\, \mapsto\, (-t,-v,-w,u)$, such 
that \eqref{ori_f} together with $v^0,f$ are anti-invariant and $v^i\,
\leftrightarrow \tilde v^i$. This involution has 
an $O3$ plane at $t=v=u=w=0$.

The transformation properties of $f(v,t|\underline{S})$ under $\sigma_1$ or $\sigma_2$
will restrict the number of allowed complex structure deformations $\underline{S}$. 
An orientifold invariant deformation in \eqref{general_deform} arises from monomials $t^{2p} v^q$ for $\sigma_1$, and 
monomials $t^{2q} v^{2p+1}$ or $t^{2q+1} v^{2p}$ for $\sigma_2$, with $p,q \in \mathbb{N}$.
We will denote the number of such monomials by $b_-$, since they 
correspond to compact three-cycles $\cA_k^-$ anti-invariant under $\sigma^*$. Respectively, the number of 
non-allowed monomials is denoted by $b_+$, and corresponds to the number of invariant compact  three-cycles
 $\cA^+_i$.
For the local Calabi-Yau $Y(A_{2k},m)$ and both involutions $\sigma_1$, $\sigma_2$ one has
\beq
   m\ \text{even:}\quad  b_\pm= b/2\ ,\qquad \qquad
   m\ \text{odd:} \quad b_\pm = \tfrac{1}{2}(b \mp k)\ ,
\eeq
where $b=m k(2k+1)$ is the total number of deformations in \eqref{general_deform}. 
The $\cN=1$ spectrum thus consists of $b_-$ complex structure deformations $S^k$
and $b_+$ $U(1)$ vectors $A^i_{\rm H}$.
Finally, in accord with \eqref{transJOmega},
the holomorphic three-form $\Omega=(du\wedge dv \wedge dt\wedge dw)/dP$, with $P$ given in 
\eqref{localssing2}, transforms with a negative sign under
the orientifold involutions.

\subsection{The visible singularity \label{delPezzogeometries}}

For the visible sector we will consider local del Pezzo 
surfaces introduced in section \ref{D7_sector}. In the following 
we will study del Pezzo singularities in compact Calabi-Yau manifolds, 
their resolutions and orientifold symmetries. In addition 
to general local considerations, we exemplify the techniques for 
an $E_8$ del Pezzo transition starting with the compact Calabi-Yau manifold 
$\bbP(18|9,6,1,1,1)$. Further examples are presented in section
\ref{eF1} and appendices~\ref{torictransitions} and \ref{mgn}.

\subsubsection{The geometry of Del Pezzo surfaces}

Let us first discuss the geometry of del Pezzo surfaces $\cB_n$ in more detail. 
This complements the analysis of section \ref{D7_sector} and prepares us for the study 
of possible orientifold involutions. Also here we will focus on the $\cB_n$
with $n=5,6,7,8$ with associated Lie groups $\mathfrak{g}=\hat D_5,\hat
E_6,\hat E_7,\hat E_8$.\footnote{Here the corresponding Lie algebras    
are denote with a hat in order to distinguish these singularities from 
the A-D-E singularities in Section \ref{singularitieswithsmallresolutions}.}
These del Pezzos are realized as hypersurfaces or complete 
intersections in weighted projective space. 
Moreover, they are elliptically fibered, with generic 
elliptic fiber realized as hypersurfaces or complete 
intersection. Let us denote
by $\bbP(d_1,\ldots,d_r|w_0,\ldots,w_m)$ the  complete intersection of
$r$ hypersurfaces of degree $d_1,\ldots,d_r$ in weighted projective 
space with weights  $w_0,\ldots,w_m$. The del Pezzo surfaces and
their generic elliptic fiber are listed in Table
\ref{tab-delPinproj_space}.

\begin{table}[h]
\begin{equation}
\begin{array}{|l|l |l |}
\hline
 \rule[-0.2cm]{0cm}{0.65cm}\ \mathfrak{g}\hspace*{.4cm}   & \ \ {\rm del\ Pezzo}\qquad\qquad\qquad & \ \ {\rm elliptic\ fibre}\\ 
\hline
\rule[-0.2cm]{0cm}{0.7cm}\ \hat D_5 \ & \ \ \bbP(2,2|1,1,1,1,1) &\ \  \bbP(2,2|1,1,1,1)\\
\rule[-0.2cm]{0cm}{0.7cm}\ \hat E_6 \ & \ \ \bbP(3|1,1,1,1) &\ \   \bbP(3|1,1,1)  \\
\rule[-0.2cm]{0cm}{0.7cm}\ \hat E_7 \ & \ \ \bbP(4|2,1,1,1) & \ \  \bbP(4|2,1,1) \\
\rule[-0.2cm]{0cm}{0.7cm} \ \hat E_8  &\ \ \bbP(6|3,2,1,1)  & \ \ \bbP(6|3,2,1)  \\
\hline
\end{array}
\label{delpezzo}
\end{equation}
\caption{\small \label{tab-delPinproj_space}
\textit{The D-E del Pezzo surfaces and their generic elliptic fibers.}}
\end{table}

As in sections \ref{simple_node} and  \ref{singularitieswithsmallresolutions} we can count the 
number of complex structure deformations associated to each generic del Pezzo 
singularity.
Table \ref{tab-delPinproj_space} allows us to specify a local equation with 
the singularity at the origin as displayed in Table \ref{tab-LocaldelpezzoSing}.
One infers that the dimension of the complex deformation spaces for del Pezzo 
surfaces $\cB_n$ with $n\ge 5$ is ${\rm dim}\, H^1(T\cB_n)=2n-8$. 
This can be checked, for example, for the $\hat E_7$ del Pezzo by counting
$3+(4-2)+(5-1)=6$ relevant monomials in $g_2,g_3,g_4$ respectively.
For the $\hat D_5$ case one needs to consider vector polynomials. 

\begin{table}[h]
\begin{center}

\begin{tabular}{| c | c |} \hline
  \hspace*{.2cm} $\mathfrak{g}$\hspace*{.4cm} &
   {\small del Pezzo geometry}  \\ \hline
$\hat D_5$  & $\begin{array}{rl}  \rule[-0.2cm]{0cm}{0.7cm}  y^2&= x^2+x e_1(w_1,w_2,z)+ f_2(w_1,w_2) \\
                             \rule[-0.2cm]{0cm}{0.7cm}  y^2&= z^2+ z g_1(w_1,w_2,x)+  h_2(w_1,w_2) \end{array}$ \\[2 mm] 
 \rule[-0.2cm]{0cm}{0.7cm}  $\hat E_6$  & \hspace*{.2cm} $y^3=x^3+x y g_1({w_1,w_2})+ x f_2({w_1,w_2})+y g_2({w_1,w_2})+g_3({w_1,w_2})$ \hspace*{.2cm} \\[2 mm] 
 \rule[-0.2cm]{0cm}{0.7cm}  $\hat E_7$  & $y^2=x^4+x^2 g_2({w_1,w_2})+x g_3({w_1,w_2})+ g_4(w_1,w_2)$ \\[2 mm]
 \rule[-0.4cm]{0cm}{0.7cm}  $\hat E_8$  & $y^2=x^3+x g_4({w_1,w_2})+ g_6(w_1,w_2)$
 \\ \hline
\end{tabular}
\caption{\small \label{tab-LocaldelpezzoSing}
\textit{The D-E del Pezzo surfaces.}}
\end{center}
\end{table}

\subsubsection{The orientifold action on del Pezzo surfaces \label{delPezzo_ori}}

On a compact Calabi-Yau space containing a del Pezzo surface, 
we are looking for a holomorphic involution $\sigma$ obeying \eqref{transJOmega}.  
In general, such involutions might not map the del Pezzo surfaces 
to its own homology class, but rather induce a more complicated action. 
This implies that $b^2_- \neq 0$, since there exist positive and negative
linear combinations of homologically non-trivial four-cycles. Examples of this type have been 
discussed e.g.~in refs.~\cite{Diaconescu:2005pc}. Here we are interested in involutions $\sigma$ preserving 
the non-trivial del Pezzo divisor in the compact space, but exclude the
trivial case for which the del Pezzo is entirely in the fix-point set of $\sigma$. 
Via the proper transform $\sigma$ acts uniquely on the coordinates of the del Pezzo 
surface. We are particularly interested in the action of the 
involutions on $H_2(S,\mathbb{Z})$ of the del Pezzo 
surface, which we call here generically $S$. Ideally we like to 
find invariant elements. Later on, for explicit compact examples, one needs to show that these 
are in the same homology class as the 2-cycles of the hidden singularity such that we can apply the 
mechanism of section~\ref{nonKaehler_res} to obtain a non-K\"ahler space.  

Let us work out the action on $H_2(S,\mathbb{Z})$ using the Lefschetz fixpoint theorem. 
The latter states that for an automorphism $\sigma$ on any $S$ the Lefschetz number $\Lambda_\sigma$
equals the Euler number $\chi(S^\sigma)$ of the fixpoint set $S^\sigma$, i.e.
\begin{equation}  \label{Lefshetz}
\chi(S^\sigma) = \Lambda_\sigma \equiv \sum_{k\ge 0} {\rm Tr} (\sigma^*(H_k
(S,\mathbb{Z}))\ .
\end{equation} 
We can use this to determine the number of $\pm $ eigenvalues of
$\sigma^*$ on $H_2(S,\mathbb{Z})$. Clearly $H_0(S)$ and $H_4(S)$ 
are invariant and the non-trivial information comes from $H_2(S)=H_{(1,1)}(S)$. 
It follows that the number of positive and the negative eigenvalues on middle 
homology are  
\begin{equation}
b_2^+=\frac{n-1+\chi(S^\sigma)}{2}, \qquad  b_2^-= \frac{3+n-\chi(S^\sigma)}{2}
\end{equation}

Let us start with the classification of involutions. A theorem of 
\cite{BB} lists all  pairs of  minimal involutions $(S,\sigma)$,
where $S$ is birational to $\mathbb{P}^2$. Lemma 1.1 of \cite{BB} states 
that minimality is equivalent to $\sigma(E)\neq E$ and $E\cap \sigma(E)\neq \emptyset$ 
for any exceptional curve $E$. In addition there are various non-minimal
involutions on each del Pezzo surface.\footnote{It turned out that 
  automorphisms of del Pezzo surfaces have been recently investigated more
  extensively in the mathematical literature \cite{Dolgachev}.}

\begin{itemize}
\item \textit{$\hat E_8$ del Pezzo:} 

Apart form an action on rational ruled surfaces one finds on the $\hat E_8$ del
Pezzo  the  famous {\sl Bertini involution} $y\mapsto -y$.
The fixpoint set is a curve of genus $g=4$
given by $\bbP(6|2,1,1)$ in the last three coordinates and the point $\bbP(6|3,2)$ so
$\chi(S^\sigma)=1+2-2g=-5$ and $b_2^-=8$. $K$ is invariant and, in fact, $-2K$ defines the
hypersurface in $\bbP(3,2,1,1)$, such that that all of the $E_8$ lattice is mapped to 
$-E_8$. On the classes one has $e_i\mapsto -2 K - e_i$ and $l\mapsto -l - 6K$. 
One can also see that the 240 rational curves Table \ref{tab:(0,1,1)} 
are reflected  on the middle configuration. The  Bertini
involution is the only minimal involution on the $\hat E_8$ del
Pezzo.\footnote{The Bertini involution has
  striking analogy with the corresponding involution on the Enriques surface
  used in the orientifold model of ref.~\cite{Grimm:2007xm}.}

As a non-minimal involution we can act by $w_1\mapsto - w_1$. The fix
point set is a curve of genus $g=1$ given by $\bbP(6|3,2,1)$ and three points 
$\bbP(6|2,1)$. Hence $b_2^+=5,\,b_2^-=4$. 

\item \textit{$\hat E_7$ del Pezzo:} 

Very similar to the Bertini involution is the {\sl Geiser involution} on the $\hat E_7$ del Pezzo   
which also acts like $y\mapsto -y$. The fixpoint set is a curve of genus 
$g=3$ given by $\bbP(4|1,1,1)$ in the last three coordinates. 
This implies that $b_2^-=7$, and  since $-K$ is invariant, we 
have $E_7\mapsto-E_7$, i.e.~$e_i\mapsto - K - e_i$ and $l\mapsto -l - 3K$. 
Again the $7,21,21,7$ rational curves in Table \ref{tab:(0,1,1)}, 
are reflected into themselves. The  Geiser
involution is the only minimal involution on the $E_7$ del Pezzo.

A non-minimal involution is obtained by the action $w_1\mapsto - w_1$. It
yields a genus $g=1$ curve $\bbP(4|2,1,1)$ as fixpoint locus,  and two points
$\bbP(4|2,1)$. This yields $b_2^+=4$ and $b_2^-=4$.           

\item \textit{$\hat E_6$ del Pezzo:} 

Also on the $\hat E_6$ del Pezzo we can act with $y\mapsto - y$. However, there 
are no minimal involutions on this del Pezzo. 
The configuration $P=f_3(x,w_1,w_2)+ y^2 f_1(x,w_1,w_2)=0$, allowing this
automorphism, is still generically smooth. By Bertinis theorem the 
non-vanishing of $dP$ outside $(x:y:w_1:w_2)=(0:0:0:0)$ has to be checked 
only on the base locus, which is given by all possible unions of the 
coordinate hyperplanes $x_i=0$. The fixpoint locus is a genus $g=1$ curve
$\mathbb{P}(3|1,1,1)$ in the last three coordinates and the point $(1:0:0:0)$, 
which also lies on  $P=0$. Hence $b_2^+=3$ and $b_2^-=4$. 

A second automorphism on the $E_6$ del Pezzo is given by  
$(y\mapsto - y,x\mapsto - x)$. The invariant equation
$P=g_3(w_1,w_2)+x^2  f_1(w_1,w_2)+ x y g_1(w_1,w_2)+  y^2  h_1(w_1,w_2)=0$
is likewise smooth as seen by checking transversality on the 
base locus. The fix point locus is given by $\mathbb{P}(3|1,1)$ in the first 
two coordinates, i.e.~3 points, and $w_1=w_2=0$,  which is a 
projective line $(y:x)$. Hence $b_2^+=5$ and $b_2^-=2$.     
 
\item \textit{$\hat D_5$ del Pezzo:} 

On the $\hat D_5$ del Pezzo there are no minimal involutions. However,
we can act by $y\mapsto -y $ which has as a fixpoint 
set only the $g=1$ curve $\mathbb{P}(2,2|1,1,1,1)$ in the last four 
coordinates, i.e.~$b_2^+=2$ and $b_2^-=4$. The configuration is 
smooth if not all $2\times 2$ minors of $\frac{\partial P_i}{\partial x_j}$ 
vanish on the base locus  unless $(x:y:z:w_1:w_2)=(0:0:0:0:0)$. This is 
the case, if we chose for $P_1=\sum_{i=1}^5 a_i x_i^2$ and 
$P_2=\sum_{j=1}^5 b_j x_j^2$ with generic coefficients $a_i,b_j$

A second involution acts as  $(y\mapsto -y,x\mapsto -x)$. 
In this case we get the fixpoint locus $\mathbb{P}(2,2|1,1,1)$ in the last three 
coordinates, which are $4$ points. We conclude then that  $b_2^+=4$ 
and $b_2^-=2$.
\end{itemize}

\subsubsection{Creating a del Pezzo singularity \label{creatingdelP}} 

In order to realize a del Pezzo in a compact Calabi-Yau manifold we 
study a del Pezzo transition. Similar to the simple transitions of section
\ref{simple_node} one starts by fixing a number of complex structure deformations to 
generate a del Pezzo singularity. There exist whole chains of such
transitions, which are best understood in a toric framework as discussed in
appendices \ref{torictransitions} and \ref{mgn}. Here we will
study only one particular example \cite{Morrison:1996pp}, which is of relevance to the large
volume compactifications of \cite{Balasubramanian:2005zx}. 

Let us start with the an elliptic fibration over $\mathbb{P}^2$ which is 
represented as a degree $18$ hypersurface in the weighted 
projective space $\bbP(9,6,1,1,1)$ as \cite{Candelas:1993dm}
\begin{equation} 
y^2=x^3+x f_{12}({\underline w})+ f_{18}({\underline w})\ ,
\label{ME_80}
\end{equation} 
where $({\underline x})=(y,x,w_1,w_2,w_3)$ are the weighted coordinates. We 
can count the complex structure deformations modulo automorphism of the ambient 
space by enumerating the elements in $\cR=\mathbb{C}^{\underline w}
[{\underline x}]/\cJ$, where $\mathbb{C}^{\underline w}
[{\underline x}]$ is the ring generated by all monomials in $\underline x$ 
and the left ideal $\cJ$ is generated by $\frac{\partial P}{\partial x_i}$, 
$i=1,\ldots,5$ with $P=y^2-x^3-x f_{12}({\underline w})- f_{18}({\underline w})$. 
We see that $y^2$ is in the ideal, $x^3$ is modulo the ideal equivalent 
to  $x f_{12}$ with $91$ monomials and from the $190$ monomials in 
$f_{18}$ the $9$ of the form $w_i^{18},w_i^{17}w_j$ are modulo the ideal 
equivalent to such with lower powers in $w_i$. This yields $h^{(2,1)}=190+91-9=272$ 
complex structure deformations, which can be parameterized by the 
complex coefficients of the independent generators of ${\cal R}$. 
Note that different to the powers $w_i^{17}w_j$, but similar to the 
powers $y^1,x^3$ we write the powers of $w_i^{18}$ with coefficient one  
to keep $P$ transversal. 

To force near the point $p=(0,0,0,0,1)$ inside  
$\bbP(18|9,6,1,1,1)$  a singularity of the form 
$\hat E_8$ in table \ref{delpezzo} we need to set the coefficients of the homogeneous 
of degree $12$ in ${\underline w}$ monomials $x w_1^m w_2^n w_3^k$    
with $9\le k \le 12 $ to zero. This yields $10$ conditions. Further we
have to set the coefficients of the homogeneous degree $18$ monomials
$w_1^m w_2^n w_3^k$ in $g_{18}$ for $13\le k \le 18$ to zero. Naively 
these are $21$ terms. However the 2 terms occurring for $w_3^{17}$ are 
already set to zero as they are not independent modulo the ideal. 
The coefficient of the term $w_i^{18}$ is constant in the standard 
parameterization of (\ref{ME_80}). We have to use a non standard 
parameterization for it to set it to zero. It can be easily shown that 
the elements of the ring that we are counting are still independent. 
So in total we can force the singularity by imposing $10+21-2=29$ 
conditions on the complex structure parameters.  

In the next step one has to resolve the del Pezzo singularity by a finite-size
del Pezzo surface.  The difference in the resolution is that, unlike the 
surgery in codimension two encountered in sections \ref{simple_node} and   
\ref{singularitieswithsmallresolutions}, here we blow up a divisor and the 
triviality of the canonical bundle of the resolved space has to be checked 
explicitly.  

\subsubsection{Triviality of the canonical bundle in the resolution}

Toric geometry is the standard and most general procedure to blow up the 
singularity at $p$ and to check that the canonical class remains trivial. 
Let us first discuss a simpler class of examples, which includes the
one of section \ref{creatingdelP}, and consider a 
Calabi-Yau space $M$ obtained as $m$-fold cover of a Fano threefold 
$D$~\cite{HirzebruchWerner}. Mori and Mukais classification list gives 
plenty of choices for $D$~\cite{MoriMukai}. In an affine coordinate patch 
parameterized by $y,x$ and $\underline{w}$ the defining equation of $M$ is of the form
\beq \label{general_Fano}
  y^m = g(x,\underline{w})\ .
\eeq
We choose the complex structure parameters in $g(x,w)$ so that a del Pezzo 
singularity of the form $\hat E_6$, $\hat E_7$ or $\hat E_8$ occurs. 

Now the blow up can be described as follows~\cite{HirzebruchWerner}. 
The branch locus $B$ is defined as the zero locus of $g(x,\underline{w})$. 
Blowing up the singular point $P$ at $x=\underline{w}=0$ in $B$ yields a complex
two-dimensional exceptional divisor $E_D=\pi^{-1}(P)$ in the smooth Fano threefold $\hat D$. 
Here $\pi: \hat D \rightarrow D_s$ is the resolution map from the smooth 
to the singular variety\footnote{See also the discussion of \eqref{res_map} for the quintic.}.  
The multiple covering of the exceptional divisor branched at the 
proper transform $\tilde B$ of $B$ on $E_D$ yields the blown-up surface $E_M$ in the 
resolved Calabi-Yau threefold $\hat M$.  
As always the blow-up is local and holomorphic and does not change the complex structure. 
However, the exceptional divisor $E_M$ in $\hat M$ is of codimension one and hence 
the canonical class $K$ can be modified. If $\varphi:\hat M\rightarrow \hat D$ 
is the $m$-fold covering map, then the canonical class of the resolved 
manifold is
\begin{equation}
K_{\hat M}=\varphi^*\big(K_{\hat  D}+\tfrac{m-1}{m}\tilde B \big)\ .
\label{canonicalresolution}
\end{equation} 
The calculation of $K_{\hat D}$ is easily done using the local geometry. The proper 
transform $\tilde B$ is given by $\tilde B=\pi^* B-{\rm ord}_P\ B\cdot E_D$, where 
$\rm ord$ is the order of vanishing of $g$ in \eqref{general_Fano}. 
Using \eqref{canonicalresolution} the condition $K_{\tilde X}=0$ can 
be ensured in explicit examples. 

In particular, in the transition of section \ref{creatingdelP} we have 
$m=2$ as seen from \eqref{ME_80}. The blow up corresponds to a $\hat E_8$ del
Pezzo and $B$ is the projective bundle $B=\mathbb{P}({\cal O}_{\mathbb{P}^2}\oplus {\cal O}_{\mathbb{P}^2}(-6))$.     
The upshot of the more general toric argument is that 
we blow up one divisor $D_E$ in $T_\Delta$ by adding a point to the polyhedron 
$\Delta$. The new $\Delta$ is still reflexive and the proper 
transform of the hypersurface constraint is the anti-canonical 
class in it. This implies that the canonical class of the resolved 
space is trivial which immediately generalizes to the complete chain 
of del Pezzo transitions.

\subsection{Numerical changes of the Euler characteristic \label{numericalchange}}

The change of the  Euler characteristic $\Delta \chi$ in the transition $M \rightarrow \hat M$ 
can be understand completely locally in terms of the singularity and in particular 
its Milnor number and the Euler number of the exceptional divisor $E_M$ in $\hat M$. 
Let us discuss this  here more systematically for the singularities  encountered so far.  
For quasihomogenous surface singularities $f=0$ the Milnor number was given 
in ref.~\cite{saito} as $\mu={\rm dim} {\cal O}/J_f$. 
If the leading terms are fermat, i.e.~of the form $\sum_{i=1}^k x_i^{n_i}$,
the Milnor number is given by
\begin{equation} 
\mu=\prod_{i=1}^k (n_i-1)\ .
\label{milnor} 
\end{equation} 
The change of the Euler characteristic is determined as follows. 
Taking the singular point out of $X$ the Euler characteristic 
changes by $\Delta \chi=\mu-1$, while gluing in the exceptional 
divisor we get \cite{HirzebruchWerner}
\beq
  \Delta \chi=\mu-1+\chi(E_X)\ .
\eeq

The change $\Delta \chi$ is easily determined for the examples we considered so 
far. For the node (\ref{conifold}) one uses \eqref{milnor} to derive $\mu=1$. So every blow-up 
increases the Euler characteristic by $\chi(\mathbb{P}^1)=2$. One can also consider 
the blow-up of the A-D-E surface singularities in Table \ref{tab-LocalSing}. The 
Milnor number is $r=\text{rank}(\mathfrak{g})$.  In total, for
the singular threefold (\ref{smallade3fold}) one derives $\mu=r (m h-1)$ using \eqref{milnor}.
The resolving manifold is a $\mathbb{P}^1$-tree forming 
the Dynkin diagram of $\mathfrak{g}$, its Euler number is $2\cdot r- (r-1)=r+1$, 
where $r-1$ comes from subtracting the intersection points, hence
$\Delta \chi=\mu-1+(r+1)=m h r$. For the $A_r$ singularity we find 
by using Table \ref{tab-LocalSing} that $\Delta \chi = m (r+1)r$. Indeed,
$ \Delta \chi/2$ is precisely the number of deformations of the $A_r$ singularity 
in \eqref{general_deform} as expected by geometric transition.

Most importantly, we can determine $\Delta \chi$ for the visible sector 
del Pezzo surfaces $\cB_n$ listed in (\ref{delpezzo}).
For the $\hat E_6,\hat E_7,\hat E_8$ del Pezzos  
we calculate  using \eqref{delpezzo} and  \eqref{milnor} that
$\mu=16,27,50$. The Milnor number for the $\hat D_5$ del Pezzo 
is $\mu=9$. Hence, given the Euler number $\chi(\cB_n)=3 +n$, we get
$\Delta \chi=16,24,36,60$ for $\cB_n,\, n>4$. In general, we can write 
\begin{equation} 
\Delta \chi= 2 h(\mathfrak{g}), 
\end{equation} 
where $h$ is the dual Coxeter number of the algebra $\mathfrak{g}$. For the relevant cases 
$h(\mathfrak{g})$ is listed in Table \ref{tab-LocalSing}.  We can now make use of 
$\Delta \chi$ to determine the rank of the map $\Pi$ given in \eqref{PiCY} which 
specifies the embedding of the del Pezzo surface $\cB_n$ into a Calabi-Yau manifold.
One first has to count the number of complex structure deformations $\Delta h^{(2,1)}$ 
which are fixed by specializing to a del Pezzo singularity such as the ones of
Table \ref{tab-LocaldelpezzoSing}.
The rank of $\Pi$ is then simply given by 
$\text{rank} (\Pi)=  \tfrac{1}{2}\Delta \chi -\Delta h^{(2,1)}$.

\section{Global Engineering of $U(1)$ Mediation \label{eF1}}

In this section we explicitly construct orientifolds of compact internal manifolds
which admit the desired properties to allow for $U(1)$ mediation of supersymmetry breaking.
We illustrate the general strategy by analyzing an explicit compact example permitting a visible 
$E_6$ del Pezzo surface and a hidden singularity with $S^3$'s arising from deformations 
of conifold singularities. Many checks of the geometric requirements are best analyzed 
in a toric realization which we will provide in appendix \ref{K3E6}. It should be noted that 
similar examples can be constructed straightforwardly within the toric framework as we show in appendix \ref{E7example}.

Let us first summarize the steps required in realizing the geometrical outset for scenarios of $U(1)$ mediation:
\begin{itemize} 
  \item[\textit{(1)}] Find a \textit{Calabi-Yau manifold with the desired singularities} for the hidden and visible sector. 
           Ensure that the visible singularity can be resolved by, for example, a del Pezzo surface $S$ as 
           described in section \ref{D7_sector} and \ref{delPezzogeometries}. 
           The hidden singularity can be of A-D-E type, and should contain conifolds which can be either resolved by 
           two-spheres or deformed by three-spheres as discussed in sections \ref{hiddenSusy} 
           and \ref{singularitieswithsmallresolutions}. 
  \item[\textit{(2)}] There should be \textit{topological relations} between
           cycles in the hidden and visible sector. These topological 
           relations can analyzed if the hidden and visible singularities are resolved by two-spheres and the del Pezzo surface 
           respectively. More precisely, some of the two-cycles in the del Pezzo $S$ should be non-trivial in the 
           Calabi-Yau space and homologous to the hidden two-spheres. Upon
           geometric transition of some of the hidden two-spheres to three-spheres as
           in section \ref{nonKaehler_res}, there are three- and four-chains connecting the
           hidden and visible sector.
  \item[\textit{(3)}] One needs to find an appropriate  \textit{geometrical orientifold symmetry} $\sigma$ on the internal space. 
           $\sigma$ has to keep the del Pezzo class invariant. As discussed in section \ref{delPezzo_ori}, there 
           will be always $\sigma$-positive and negative two-cycles in $S$ if $\sigma$ is non-trivial 
           on~$S$. One of the positive two-cycles in $S$ should be non-trivial in the Calabi-Yau space and in 
           topological relation to the hidden two-spheres. Some of the hidden two-spheres are exchanged under $\sigma$
           such that hidden $U(1)$ vectors arising upon geometric transition remain in the spectrum. Upon integrating 
           out the massive moduli there will be light $U(1)$ vector multiplets coupling to both the hidden and visible sector
           as discussed in section \ref{U(1)_string} and \ref{D7_sector}.
  \item[\textit{(4)}] \textit{Local engineering of the background fluxes and gauge groups.} To make such a scenario 
          fully realistic one needs to place appropriate branes on $S$ to get an extension of the MSSM.
          In the hidden sector one performs a geometric transition of the resolving two-spheres and places 
          appropriate fluxes on the resulting three-cycles to break supersymmetry. It has to be ensured 
          that the scales are appropriate to yield an interesting set of soft terms for the MSSM sector. 
\end{itemize}

In the following we will realize steps $(1)$, $(2)$ and $(3)$ for one compact 
example. The geometric outset for two further examples are presented in appendices \ref{K316con} and \ref{E7example}.
Our first investigation will start from the compact Calabi-Yau obtained 
from a singular quintic as considered in ref.~\cite{Malyshev}. 
It is realized via the complete intersection 
\bea \label{Compl_Int_1}
  \big(P_1 s + P_2 t \big)\, u +  \big(Q_1 s + Q_2 t \big)\, v &=&0 \ , \nn \\
  \big(R_2 s + R_3 t \big)\, u +  \big(S_2 s + S_3 t \big)\, v &=&0 \ ,
\eea
in an ambient space defined as $\bbP^1 \times \bbP(\cO_{\bbP^3}(-1) \oplus \cO_{\bbP^3})$. Here $P_i,Q_i,R_i$ and $S_i$
are monomials of degree $i$ in the complex projective 
coordinates $\underline{w}=(w_1,w_2,w_3,w_4)$ on the $\bbP^3$ base,
while the coordinates on the $\bbP^1$ fibers are denoted by $s,t$. The $\bbP^1$ factor has 
coordinates $u,v$.  The toric data of this Calabi-Yau manifold $Y_{E_6}$ are summarized in appendix \ref{K3E6}.
$Y_{E_6}$ is a K3 fibration  with 
\beq \label{Hodge_E6dP}
   h^{(1,1)}=3\ ,\qquad h^{(2,1)}=59\ ,\qquad [c_2]_1 = 44\ ,\qquad [c_2]_2 = 50\ ,\qquad [c_2]_3 = 24\ .
\eeq
where $[c_2]_i=\int_{Y_{E_6}}  c_2 \wedge  \omega_i$ with $\omega_i$ being a basis of $H^2(Y,\bbZ)$.
The non-vanishing triple intersections $\kappa_{ijk} = \int_{Y_{E_6}} \omega_i \wedge \omega_j \wedge \omega_k$ 
are computed to be
\begin{equation}  \label{Intnumbers_E6dP}
\kappa_{111}= 2, \qquad
\kappa_{112}=\kappa_{122}=\kappa_{222}=5, \qquad
\kappa_{113}=4, \qquad
\kappa_{123}=\kappa_{223}= 6\ .
\end{equation} 

In the following we will check that $Y_{E_6}$ admits in its moduli space two singularities 
of the desired type. This includes an $\hat E_6$ del Pezzo at $(s,t,u,v,\underline w)=(1,0,0,0,\underline{0})$
as well as $32$ conifold singularities.
To see this one considers a point where at least one of the functions in front of $u,v$ is non-zero and eliminates
$u,v$ in \eqref{Compl_Int_1} to 
\beq \label{uv_elimination}
  \big(P_1 s + P_2 t \big) \big(S_2 s + S_3 t \big)- \big(R_2 s + R_3 t \big) \big(Q_1 s + Q_2 t \big)= 0
\eeq
The term in front of $s^2$ yields a non-generic cubic surface in coordinates $\underline{\omega}$ of $\bbP^3$ 
given by 
\beq \label{nongen_E6-delP}
    P_1 S_2 - R_2 Q_1= 0\ .
\eeq
{}From table \ref{tab-delPinproj_space} we identify this as a non-generic $\hat E_6$  del Pezzo surface.
To identify the $32$ resolved conifolds we compare \eqref{uv_elimination} and \eqref{Compl_Int_1} to the 
simple singular conifold \eqref{node} and its resolution by a two-sphere \eqref{res}.
We see that the singular conifolds are obtained if all four factors in 
front of $u,v$ in \eqref{Compl_Int_1} vanish simultaneously. 
In fact, generically there are $32$ conifold points where these vanishing conditions can be satisfied. 
At generic points on the moduli space $Y_{E_6}$ is smooth and the conifolds are resolved by $32$ two-spheres.
The Calabi-Yau space $Y_{E_6}$ can be obtained via geometric transition 
from a quintic surface by enforcing the $\hat E_6$ del Pezzo and $32$ conifolds by fixing $\Delta h^{(2,1)}=42$
of the $101$ quintic complex structure deformations~\cite{Malyshev}. These transitions 
are discussed torically in appendix~\ref{K3E6}. 

In the next step we need to check that there is a two-cycle in the $\hat E_6$ del 
Pezzo $\cB_6$ which is non-trivial in $Y_{E_6}$ and in the same homology class as the 
$32$ resolved conifolds. We will explicitly identify the two-cycles in 
$\cB_6$ which are non-trivial in $Y_{E_6}$.
Note that since $h^{(1,1)}=3$ there are in total three non-trivial 
two-cycles in $Y_{E_6}$, which are denoted 
by $\Sigma_B$, $\Sigma_{K}$ and $\Sigma_{\alpha}$. The two-cycle $\Sigma_{K}\cong -K$ corresponds 
to the anti-canonical class of the $\hat E_6$ del Pezzo embedded into $Y_{E_6}$. In case this is the only 
class non-trivial in $Y_{E_6}$ one will find $27$ genus zero curves in its homology class. These 
are the $27$ lines in table~\ref{tab:(0,1,1)}, which by \eqref{deg_g} have
genus zero and degree one with respect to $-K$.
In fact, we can compute the genus zero BPS invariants $n_{n,m,l}$ by using the toric description of 
$Y_{E_6}$ in appendix~\ref{K3E6}. The three indices of $n_{m,n,l}$ indicate the degree of the 
curve with respect to $\Sigma_{K}$, $\Sigma_B$ and $\Sigma_{\alpha}$. If $\Sigma_{K}$
is the only non-trivial class of $\cB_6$ in $Y_{E_6}$ one has $n_{1,0,0}=27$. However, 
one instead computes 
\beq \label{BPS_E6}
   n_{1,0,0}=10\ ,\qquad  n_{1,0,1}=16\ ,\qquad n_{1,0,2}=1\ , \qquad \qquad n_{0,0,1}=32\ .
\eeq
Since $Y_{E_6}$ can be realized torically as in appendix \ref{K3E6}
the BPS numbers can be obtained with the methods explained in
refs.~\cite{HKTY}.\footnote{
The explicit calculations are implemented in the program Instanton, which uses
the toric data as input parameters.}   
This shows that there is actually a second curve class $\Sigma_\alpha \cong \alpha$
in the del Pezzo surface which is non-trivial in $Y_{E_6}$. 
In accord with \eqref{BPS_E6} one identifies
\beq \label{alpha_inCY}
   \alpha=2l-e_1 -e_2 -e_5 - e_6\ , \qquad \quad \alpha \cdot K = - 2\ .
\eeq 
where $l,e_i$ and $K$ are introduced in section \ref{D7_sector}. Using \eqref{deg_g} one
checks that there are $10$, $16$ and $1$ curves of table \ref{tab:(0,1,1)}
with degree $0,1$ and $2$ with respect to $\alpha$. We can write this 
as ${\bf 27} = {\bf 10}_0+ {\bf 16}_1+ {\bf 1}_2$, where the subscripts correspond to the degree with respect to $\alpha$. 
This decomposition corresponds to the splitting of the ${\bf 27}$ of $E_6$ into representations of $U(1)\times D_5$.
We can shift $\alpha$ by the other non-trivial class $K$ such that the new linear combination is 
in the $E_6$ lattice. Defining $\alpha'=3 \alpha +2 K$ one has $K\cdot \alpha'=0$. The new class $\alpha'$ induces 
the splitting 
\beq
 {\bf 27} = {\bf 10}_{-2}+ {\bf 16}_1+ {\bf 1}_4\ ,
\eeq
where the subscripts indicate the degree with respect to $\alpha'$ and represent the $U(1)$ charge in the usual 
normalization  \cite{Slansky}. The last BPS invariant $n_{0,0,1}$ in \eqref{BPS_E6} shows that there 
are actually $32$ genus zero curves in the homology class of $\Sigma_{\alpha}$. These are the $32$
resolved conifolds discussed above. This establishes the desired topological relation of 
the del Pezzo to the candidate hidden singularities.

Let us now turn to the orientifold projection. As we have seen in section
\ref{delPezzo_ori}, there are two non-trivial orientifold projections on the $\hat E_6$ 
del Pezzo. One of them inverts two coordinates and thus, in order to ensure \eqref{transJOmega}, would have to be accompanied 
by an inversion of an additional coordinate in the Calabi-Yau embedding.  
In order to avoid this complication, we will focus on the second involution 
which only inverts one coordinate, say $w_4 \rightarrow - w_4$ in 
\eqref{Compl_Int_1}. In general, a $\hat E_6$ del Pezzo surface admitting this
involution can be brought to the form 
\beq
   w_4^2 (\alpha \, w_1 + \beta \, w_2 +\gamma \, w_3) + w_1^3 + w_2^3 + w_3^3
   + \delta\, \omega_1 \omega_2 \omega_3=0\ .
\eeq
This del Pezzo can be brought to the non-generic form \eqref{nongen_E6-delP}
by fixing two of the four complex structure deformations $\alpha,\beta,\gamma,\delta$. 
Slightly redefining coordinates one has
\beq
  w_1(w_1^2 - 3w_1 w_3 + \delta\, w_2 w_3 + 3 w_3^2) + w_2 (w_2^2 - \delta\,
  w_3^2 + \beta\, w_4^2) =0 \ .
\eeq

Next, we can determine a basis of $H_2^+(\cB_6) \oplus H^-_2(\cB_6)$. Recall from 
section \ref{delPezzo_ori} that $b_+^2=3$, $b_-^2=4$ for a
del Pezzo surface with the above involutive symmetry. Since the canonical class of the del Pezzo
is invariant under $\sigma$, one can embed the orientifold involution into the Weyl group of $E_6$.
$\sigma$ will be represented by four Weyl reflections $\sigma_a:\, \beta \mapsto \beta + (\beta \cdot r_a)\, r_a$ 
on four mutually orthogonal roots $r_a$ \cite{Carter}.
Up to coordinate redefinitions one finds
\beq
 \sigma^* = \prod_{a=1}^4 \sigma_a\ ,\qquad   r_1 = \alpha_{\rm max}\ , \quad r_2 = \alpha_1\ ,\quad r_3= \alpha_3\ ,\quad r_4 = \alpha_5\ ,
\eeq
where $\alpha_i$ are the simple roots defined in \eqref{delpezzolattice}, and $\alpha_{\rm max} = 2l - \sum_i e_i$ is the maximal root of $E_6$.
It is easy to check that the $\Sigma^-_a = r_a$ are a basis of $H^-_2(\cB_6)$. Such a correspondence 
is a general fact for involutions in a Weyl group \cite{Carter}. A basis of 
$H_2^+(\cB_6)$ is given by $\Sigma_3^+=-K$ as well as the two elements
\beq
   \Sigma_1^+ = 2 \alpha_2 + \alpha_1 +\alpha_3 \ ,\qquad \Sigma_2^+ = 2
   \alpha_4 + \alpha_3 + \alpha_5\ .
\eeq
Finally, we can express the element $\alpha'$ given after \eqref{alpha_inCY} in this basis,
$\alpha'= 2 \Sigma_2^+ - 2 \Sigma_1^+$. This shows that the second del Pezzo 
two cycle $\Sigma_\alpha$, which is non-trivial in the full Calabi-Yau space, is also positive under the 
orientifold involution. Hence, for the Calabi-Yau orientifold space $(Y_{E_6},\sigma)$ 
we have $h^{(1,1)}_+ = 3$, $h^{(1,1)}_-=0$. The dimensions of $h^{(2,1)}_\pm$
can now be evaluated by using the Lefschetz fixpoint theorem as in \eqref{Lefshetz}.

In order that $Y_{E_6}$ is a candidate to permit $U(1)$ mediation, 
we have to show that there are actually hidden $U(1)$ vectors in the spectrum. 
To do that, one shows that the $32$ two-spheres in the homology class of $\Sigma_\alpha$ 
are not all mapped to themselves under $\sigma$. In fact, one checks that
generically only two conifold points are invariant under $w_4 \rightarrow -
w_4$, while the remaining $30$ points are pairwise identified. This will
remain to be the case if these singularities are resolved by two-spheres or deformed by 
three-spheres.
Finally, one needs to proceed as in section \ref{nonKaehler_res} to construct the
non-K\"ahler space $\cM_6$. One performs a geometric transition replacing the
$30$ pairwise identified hidden two-spheres by three-spheres. Clearly, the
resulting three-spheres will also be identified pairwise under the involution $\sigma$,
such that several $U(1)$ vectors from the R-R four-form remain in the
spectrum.  Since the two-spheres were in topological relation with the del
Pezzo, the visible and hidden
sector will be naturally connected via three- and four-chains in the
non-K\"ahler space~$\cM_6$.

\section{Conclusions}

In this paper we investigated a promising mediation mechanism for
supersymmetry breaking due to background fluxes in Type IIB string theory. Our
motivation was the generic presences of $U(1)$ vectors in
semi-realistic string constructions of extended Supersymmetric Standard Models. We
argued that under certain topological conditions these $U(1)$ vector
multiplets can couple to a hidden supersymmetry breaking flux
geometry. Non-vanishing F-terms can render the  gauginos massive which 
then induce phenomenologically interesting soft masses in the visible sector. 

In order to find explicit realizations of such scenarios various requirements on
the underlying compactification manifolds have to be met. Firstly, it should
contain four-cycles on which intersecting space-time filling D-branes can provide
an extension of the Standard Model. Promising candidates for such
constructions are del Pezzo surfaces and we have provided a list of concrete compact 
Calabi-Yau manifolds admitting these as exceptional divisors. Secondly, the
internal manifold should admit singularities which can support a supersymmetry
breaking flux background. These are typically present in most of the known
Calabi-Yau manifolds. However, the cycles in this hidden flux geometry have to be in 
topological relation with the visible sector geometry. In sections \ref{geometric_real} and \ref{eF1}, we
have summarized the general strategy to construct viable orientifold
geometries for $U(1)$ mediation. 

Hypersurfaces and complete intersections provide a 
vast class of compact examples in which the required local geometries can be
embedded and studied in detail. Combining the counting of the rational
holomorphic curves in the compact Calabi-Yau manifold with the representation
theory of the Weyl group acting on the curves in the del Pezzo surface, we were able
to identify globally realized classes and their topological relations. 
Using the Lefschetz fixpoint theorem together with the embedding of
the orientifold involution into the Weyl group, allows
the determination of the parity of the curve classes.
This provides the ground for explicit model building in type IIB orientifolds
and F-theory. The described techniques are useful not only in modeling $U(1)$ mediation.
In particular, it should be possible to embed the geometric realizations of
dynamical supersymmetry breaking \cite{ABK} into the explicit Calabi-Yau 
orientifolds investigated in this work.  

Let us comment on some of the open questions and directions for future work.
In our investigation, we have not addressed the stabilization of moduli in detail.
For type IIB orientifolds there are various known mechanism to render all geometric
moduli massive. However, their stabilization might not entirely decouple from the low
energy phenomenology investigated here \cite{Blumenhagen:2007sm}. The constructed compact
orientifolds can, for example, support KKLT like vacua~\cite{KKLT} or the large volume vacua
studied in~\cite{Balasubramanian:2005zx}.
In particular, one can show that some of the examples realizing del Pezzo transitions
as in section \ref{geometric_real}  and \ref{eF1}  have the desired properties for the large volume scenarios.
It will be interesting to work out the explicit $\cN=1$ data for these
examples and to study the supersymmetry breaking and moduli stabilization in
more detail.

In order to get the correct scales of supersymmetry breaking it is often
crucial to also include warping effects into the set-up. In particular, a
hidden flux sector will generically break supersymmetry at a rather large
scale. It is an interesting problem to study $U(1)$ mediation of supersymmetry
breaking in the presence of strongly warped throats. For this a better understanding of
Kaluza-Klein reduction in warped backgrounds will be crucial.

\section*{Acknowledgments}

We are grateful to Ralf Blumenhagen, Tae-Won Ha, Shamit Kachru, Denis Klevers, Eduard Looijenga, Dieter L\"ust, Peter Mayr, Hans-Peter
Nilles, Kyriakos Papadodimas and Alessandro Tomasiello for useful discussions.
We in particularly thank Eduard Looijenga and Shamit Kachru for valuable 
correspondence.  This work was partially supported by the
European Union 6th framework program MRTN-CT-2004-503069
``Quest for unification", MRTN-CT-2004-005104 ``ForcesUniverse", MRTN-CT-2006-035863 
``UniverseNet" and SFB-Transregio 33 ``The Dark Universe" by the DFG.

\vspace{2cm}

\appendix

\noindent {\bf \LARGE Appendices}
\section{Toric realizations of the visible and the hidden singularities }
\label{torictransitions}

\subsection{K3 fibration with an $\hat E_6$ del Pezzo and $16$ conifolds \label{K316con}}
  
In this appendix we will discuss the principles of toric transitions 
starting with the Calabi-Yau which is realized as the quintic 
hypersurface in $\mathbb{P}^4$. This manifold allows an $\hat E_6$ del Pezzo 
transition as well a transition through a singularity, which 
can serve as the hidden sector. Moreover, for the final Calabi-Yau  
there is a topological relation between a curve class on the del 
Pezzo surface, which serves as the visible sector, and a curve 
class in the hidden sector. This topological relation can be studied by
computing the relevant BPS invariants and arguing as in section \ref{eF1}.

It turns out that such transitions are frequently realized for Calabi-Yau in
toric ambient spaces. Let us recapitulate the essentials of the 
toric construction. We list in the following table in the first column the
data of the reflexive polyhedra involved. More precisely the polyhedron
$\Delta$ which yields  $\mathbb{P}_\Delta=\mathbb{P}^4$, the ambient space of 
the {quintic}, is given by the convex hull of the first five points 
$\nu_1,\ldots,\nu_5$.  If we add the point $\nu_6$ we enforce {transition
  I}. We get as convex hull of the points $\nu_1,\ldots,\nu_6$ a polyhedron
$\Delta^{(2)}$, which defines as  $\mathbb{P}_{\Delta^{(2)}}$ the ambient space of a Calabi-Yau 
with an $\hat E_6$ del Pezzo. Similarly, we can consider the convex hull of the points
$\nu_1,\ldots ,\nu_5,\nu_7$ to enforce {transition II}. Finally, 
if we consider all 7 points $\nu_1,\ldots,\nu_7$ we are dealing with
{transition III} in the table. In the following we will discuss these
transitions in more detail.

\begin{center}
\begin{tabular}{rlrrr|rr|rrr|rrr|rrrr}
&&&&& \multicolumn{2}{c|}{{quintic}} & \multicolumn{3}{c|}{{transition I}} & 
\multicolumn{3}{c|}{{transition II}} &   \multicolumn{4}{c}{{transition III}} 
\\[.2cm]
\cline{6-17}
\rule[-0.2cm]{0cm}{0.8cm} &&&&&&$l^{(1)}$&&$l_1^{(2)}$&$l_2^{(2)}$
&&$l_1^{(3)}$& $l_2^{(3)}$& &$l_1^{(4)}$& $l_2^{(4)}$& $l_3^{(4)}$     \\
\multicolumn{5}{l|}{points in polyhedra} &$x_0$&-5&                                 & -3 &-2 &$x_0$ &-4&-1 &$x_0$&-2&-2&-1     \\
\hline 
&&&&&&&&&&&&&                        \\
$\nu_1$&=(\ -1,&0,&0,&0)& $x_1$& 1&  $x_1$      &1 &0& $u_1$    &0& 1& $u_1$&  0& 0&1                        \\
$\nu_2$&=(\ \ 0,&-1,&0,&0)&$x_2$&1&  $x_2$      &1 &0&  $v_1$   &1& 0& $v_1$&  1& 0&0                       \\ 
$\nu_3$&=(\ \ 0,&0,&-1,&0)&$x_3$&1&  $x_3$      &1 &0&  $v_2$   &1& 0& $v_2$&  1& 0&0                       \\ 
$\nu_4$&=(\ \ 0,&0,&0,&-1)&$x_4$&1&  $s$        &0 &1&  $v_3$   &1& 0& $s$ &   0& 1&0                       \\ 
$\nu_5$&=(\ \ 1,&1,&1, &1)&$x_5$&1&  $x_4$      &1 &0&  $u_2$   &0& 1& $u_2$&  0& 0&1                      \\ 
$\nu_6$&=(\ \ 0,&0,&0,&1)& &&   $t$        &-1&1&          & &  & $t$ &  -1& 1&0                      \\
$\nu_7$&=(\ \ 0,&1,&1,&1)& & &             &  & &   $r$    &1& -1&$r$ &   1&0&-1      \\
\end{tabular}
\end{center}
\vspace{.4cm}

As usual we define $\mathbb{P}_\Delta$ by the Cox coordinate ring as
$\mathbb{P}_\Delta=\{ C[x_1,\ldots,x_n]-SR\}/(\mathbb{C}^*)^r$. Here the
$\mathbb{C}^*$ actions are specified by the vectors $l^{(p)}_k$, $k=1,\ldots,r$ 
as $x_i\mapsto (\mu_{k})^{l^{(p)}_{k,i}}\ x_i$, with $\mu_{k}\in
\mathbb{C}^*$. The $SR$ is the Stanley Reisner ideal, which is also specified by the  
$l^{(p)}_k$, see \cite{Cox} for details. The index $(p)$ on $l^{(p)}_i$ 
labels the Calabi-Yau phase under consideration after the respective transition.  
Note that the $l^{(p)}_k$ encode relations 
among the points in the polyhedron namely $\sum_{i=1}^n
l^{(p)}_{k,i}\, \nu_i=0$, where $n$ is the number of relevant points. Using suitable triangulations of the polyhedron 
we have specifically chosen them so that in the Calabi-Yau phase indexed 
by $(p)$ to every $l^{(p)}_k$, $k=1,\ldots,h^{(1,1)}$ there is an
uniquely associate K\"ahler modulus $v^k$ of a curve in the Calabi-Yau, 
so that the K\"ahler cone is given by $v^k> 0$.

The hypersurfaces or complete intersections represent the anti-canonical 
class of $\mathbb{P}_\Delta$ and are easily defined by the $l^{(p)}_k$,
$k=1,\ldots,r$. In the case of hypersurfaces the polynomial $P(x)$, 
whose zero section defines the Calabi-Yau in the coordinates 
$x_1,\ldots,x_m$ is simply such that $x_0 P(x)$ is totally 
invariant under the $(\mathbb{C}^*)^r$ actions. 

Each point point of $\Delta$ corresponds to a toric divisor given as the 
zero locus of the corresponding coordinate, e.g.~$\nu_1$ 
corresponds to $x_1=0$ also called $D_1$. Among the divisors there are 
relations $\sum_{i=1}^{n}\nu_{i,k} D_i=0$, where $\nu_{i,k}$ is the $k$ 
component of $\nu_i$. The Chern class of the  ambient space is 
$c(T\mathbb{P}_\Delta)=\prod_{i=1}^{n}(1+D_i)$, the canonical 
class is $K=\sum_{i=1}^{n} D_i$. The total Chern class of the 
Calabi-Yau $M$, which is specified by $-K$, is $c(TM)= c(T\mathbb{P}_\Delta)/(1+K)$.
The $k$-th Chern class $c_k$ are then the terms homogeneous of order $k$ in
the $D_i$ in the formal expansion of $c(TM)$. So by construction $c_1(TM)=0$.

In the formalism of reflexive pairs $(\Delta,\Delta^*)$ the transitions are very easily
understood. Points in $\Delta$ count the K\"ahler parameter, while  points in
$\Delta^*$ counts complex parameters. If we add the point $\nu_6$ to enlarge
$\Delta=\Delta^{(1)}$ to $\Delta^{(2)}$, we create a del Pezzo singularity 
in the first $4$ coordinates, because the enlargement of $\Delta$ enforces the vanishing of $11$
coefficients of the monomials, which correspond to complex structure deformations in the
Newton polynom of $\Delta^*$. So $\Delta^{(2)}\supset \Delta$, which counts the K\"ahler
parameter, becomes larger while its dual $\Delta^{(2)*}\subset \Delta^*$, which
counts the complex parameters, becomes smaller.

In the table we have  chosen coordinates which are most intuitive to 
understand the precise form of the polynomial constraint after the different 
transitions.  For the quintic $P({\underline x})$ is simply given by a homogeneous polynomial 
of degree five in the first five variables. Then after the transition I
the constraint equation looks like
\begin{equation} 
P=x_0\sum_{k=0}^2 p_{3+k}({\underline x}) s^{2-k} t^k\ .
\end{equation}
Here the $p_k({\underline x})$ are homogeneous polynomials of degree $k$ in 
$x_1,\ldots, x_4$. It is obvious that at $t=0$ and $s=1$ we get a $E_6$ del
Pezzo singularity. General $s,t$ with the above $\bbC^*$ actions correspond 
to the blow up of the del Pezzo singularity. Adding
$l^{(1)}=l^{(2)}_1+l^{(2)}_2$, dropping the $t$ variable which scales
trivially under $l^{(1)}$, and identifying $s$ with $x_5$ allows to deform 
the del Pezzo singularity to a generic quintic. 

Let us analyze the Calabi-Yau obtained after transition I in more detail. 
As seen from $l^{(2)}$ the toric variety is $\mathbb{P}({\cal O}_{\mathbb{P}^3} \otimes  
{\cal O}(-K)_{\mathbb{P}^3})$.
We have the following topological data for the manifold:  
$\chi=-176,\, h^{(1,1)}=2$ and 
\begin{equation}
[c_2]_1=44\ , \quad  [c_2]_2=50\ ,\quad
\kappa_{111}=2\ ,\ \ \kappa_{112}=\kappa_{122}= \kappa_{222}=5\ ,
\end{equation} 
where $[c_2]_i = \int_M c_2 \wedge  \omega_i$ and $\kappa_{ijk}=\int_M
\omega_i \wedge \omega_j \wedge \omega_k$ as in \eqref{Hodge_E6dP} and \eqref{Intnumbers_E6dP}.
It is also instructive to list the genus zero BPS invariants for the rational
curves $n_{i,j}$, because we see here the $27$ lines of the $\hat E_6$ del Pezzo 
in the class $(1,0)$, which corresponds to the class~$-K$.   
\begin{center}
\underline{$n_{i,j}$ for trans.~I\ :} \qquad \begin{tabular}{r|rrrrrrr}
i&j=0&j=1&j=2&j=3&j=4&j=5&j=6\\
\hline
0&0&    60&     0&      0&      0&      0&      0 \\
1&\bf{27}&   2515&   12210&  12210&  2515&   27&     0\\
\hline
\end{tabular}
\end{center}

Let us next turn to the transition II listed in the table above. After this transition the polynomial is of the form
\begin{equation} 
P=x_0\sum_{k=0}^4 r^k p_{4-k}({\underline v}) q_{k+1}({\underline u}) \ .
\end{equation}
We have now a singularity at $r=0$, which is a $\mathbb{P}^1$ bundle 
over a curve of genus $3$. It has been analyzed in \cite{Candelas:1993dm,Katz:1999xq}, where it has been 
argued that it can be deformed to isolated $16$ $\mathbb{P}^1$, i.e.~$16$
conifolds.  This manifold is a $K3$ fibration with the topological data  
$\chi=-168,\, h^{(1,1)}=2$ and
\begin{equation}
 [c_2]_1=50\ , \quad  [c_2]_2=24\ ,\quad
\kappa_{111}=5\ ,\ \ \kappa_{112}=4\ .
\end{equation} 
For this Calabi-Yau manifold the BPS instantons are as follows
\begin{center}
 \underline{$n_{i,j}$ for trans.~II\ :} \qquad \begin{tabular}{r|rrrrrrr}
i&j=0&j=1&j=2&j=3&j=4&j=5&j=6\\
\hline
0&0&    \bf{16}&     0&      0&      0&      0&      0\\
1&640&  2144&   120&    -32&    3&      0&      0\\
\hline
\end{tabular}
\end{center}
In particular we see that $n_{0,1}=16$ which corresponds to the
configuration which can be deformed to $16$ conifolds.

Finally, we turn to the transition III for which both points 
$\nu_6,\nu_7$ are added. This leads to a polynomial of the form
\begin{equation}
P=x_0 \sum_{k=0}^2 t^k s^{2-k} \sum_{l=0}^{k+2} r^l p_{2+k-l}({\underline u})
q_{l+1}({\underline v}) \ . 
\end{equation} 
We see that at $t=0$, $s=1$ there is a non-generic del Pezzo singularity, 
while at $r=0$ there is  a degenerate version of the $\mathbb{P}^1$ bundle 
over a curve of genus $3$, which can be deformed to $12$ isolated 
$\mathbb{P}^1$'s, i.e.~$12$ conifolds.
Again this manifold is a $K3$ fibrations with 
\begin{equation}
\chi=-152\ ,\quad  h^{(1,1)}=3\ ,\quad [c_2]_1=24, \quad  [c_2]_2=50\ ,\quad
[c_2]_3=44\ , 
\end{equation} 
and the non-vanishing triple intersections      
\begin{equation} 
\kappa_{111}=\kappa_{113}= 2\ , \ \ 
\kappa_{112}=\kappa_{122}=\kappa_{222}=5\ , \ \ 
\kappa_{123}=\kappa_{223}=4\ .
\end{equation} 
The genus zero BPS instantons $n_{0,i,j}$, $n_{1,i,j}$ and $n_{2,i,j}$  are given by 
\begin{center}
\begin{minipage}{14cm}
\underline{$n_{0,i,j}$ for trans.~III\ :} \qquad \begin{tabular}{r|rrrrrrr}
i&j=0&j=1&j=2&j=3&j=4&j=5&j=6\\
\hline
0& 0&   \bf{12}&     0&      0&      0&      0&      0\\
0&50&   12&     -2&     0&      0&      0&      0\\
\hline
\end{tabular}
\vspace*{.4cm}

\underline{$n_{1,i,j}$ for trans.~III\ :} \qquad \begin{tabular}{r|rrrrrrr}
i&j=0&j=1&j=2&j=3&j=4&j=5&j=6\\
\hline
0& \bf{10}&   \bf{16}&     \bf{1}&      0&      0&      0&      0\\
1&540&  1920&   76&     -24&    3&      0&      0\\
2&1396& 10064&  1035&   -440&   198&    -48&    0\\
\hline
\end{tabular}

\end{minipage}
\end{center}

\subsection{K3 fibration with $\hat E_6$ del Pezzo and 32 conifolds \label{K3E6}}

In this appendix we present the toric analysis of the Calabi-Yau space 
used in section~\ref{eF1}. The geometric construction of this example 
has been discussed by Malyshev in ref.~\cite{Malyshev}. Our investigation will follow a 
similar logic as in appendix \ref{K316con} and we will likewise start with the quintic in
$\mathbb{P}^4$. Again in the table polyhedron of $\mathbb{P}^4$ is the four 
simplex given the convex hull by the points $\nu_1,\ldots,\nu_5$ in 
$\mathbb{R}^4$, i.e.~with the last entry dropped. The transition I to 
the Calabi-Yau with the $\hat E_6$ del Pezzo singularity is exactly as in appendix 
\ref{K316con}.  
Since we like to  study latter a transition to a complete intersection 
in a five dimensional space given by a five dimensional polyhedron we 
added a trivial fifths coordinate to the first six  points. 

\begin{center}
\begin{tabular}{rlrrrr|rr|rr|rr|rrr}
&&&&&& \multicolumn{2}{c|}{{quintic}} & \multicolumn{2}{c|}{{trans.~I}} & 
\multicolumn{2}{c|}{{trans.~II}} &   \multicolumn{3}{c}{{trans.~III}} 
\\[.2cm]
\cline{7-15}
\rule[-0.2cm]{0cm}{0.8cm} &&&&&&&$l^{(1)}$&$l_1^{(2)}$&$l_2^{(2)}$&$l_1^{(3)}$& $l_2^{(3)}$&$l_1^{(4)}$& $l_2^{(4)}$& $l_3^{(4)}$     \\
&&&&&&$x_0'$                               &  &  &   &-3&-1 &-2&-1&-1     \\
&&&&&&$x_0$      &-5&-3&-2 &-2&-1 &-1&-1&-1     \\
\hline 
&&&&&&&&&&&&&                        \\
$\nu_1$&=(\ -1,&0,&0,&0& 0)& $w_1$       &1 &1 &0&1 &0&    1& 0&0                        \\
$\nu_2$&=(\ \ 0,&-1,&0,&0&0)&$w_2$         &1&1 &0&1 &0&    1& 0&0                       \\ 
$\nu_3$&=(\ \ 0,&0,&-1,&0&0)&$w_3$         &1&1 &0&1 &0&    1& 0&0                       \\ 
$\nu_4$&=(\ \ 0,&0,&0,&-1&0)&$s$         &1&0 &1&1 &0&     0& 1&0                       \\ 
$\nu_5$&=(\ \ 1,&1,&1, &1&0)&$w_4$         &1&1 &0&1 &0&    1& 0&0                      \\ 
$\nu_6$&=(\ \ 0,&0,&0,&1&0)&$t$       & &-1&1&  & &   -1& 1&0                      \\
$\nu_7$&=(\ \ 0,&0,&0,&0&1)&$u$        & &  &  &0 &1&    0& 0&1\\
$\nu_8$&=(\ \ 0,&0,&0,&0&-1)&$v$        &&  &  &0 &1&    0& 0&1\\
\end{tabular}
\end{center}

Let us turn to the transition II. The $l^{(3)}$ are the scalings for a
conifold transitions after generating $36$ nodes.  
The two polynomials $P$ and $P'$ are defined by the $l^{(3)}$  
requiring that $x_0P $ and $x_0'P$ are invariant under the $(\mathbb{C}^*)^2$ 
actions. The same is true for the transition III. 
We have the following topological data for the manifold after the transition II.
It is a $K3$ fibration with $\chi=-128,\, h^{(1,1)}=2$ and
\begin{equation}
[c_2]_1=50\ , \quad  [c_2]_2=24\ ,\quad
\kappa_{111}=5\ ,\ \ \kappa_{112}=6\ , \ \ \kappa_{122}= \kappa_{222}=0\ ,
\end{equation} 
with a notation as in \eqref{Hodge_E6dP} and \eqref{Intnumbers_E6dP}.
From the genus zero BPS invariants for the rational we see the homologous
$36$ $\mathbb{P}^1$ blow-ups of the nodes in the class $(0,1)$ as expected.    
\begin{center}
\underline{$n_{i,j}$ for trans.~II\ :} \qquad \begin{tabular}{r|rrrrrrr}
i&j=0&j=1&j=2&j=3&j=4&j=5&j=6\\
\hline
0&0&   \bf{36}&     0&      0&      0&      0&      0\\
1&366&  1584&   909&    16&     0&      0&      0\\
2&2670& 73728&  255960& 231336& 45216&  360&    -20\\
\hline
\end{tabular}
\end{center}

Finally, we can analyze the transition III. The resulting Calabi-Yau manifold
and with a candidate orientifold projection has been studied in section \ref{eF1}.
In the toric set-up we now have the scalings $l^{(4)}$ which 
correspond to performing both transitions simultaneously. We get a $K3$ 
fibration with  topological data as in \eqref{Hodge_E6dP} and 
non-vanishing triple intersections \eqref{Intnumbers_E6dP}.  
The BPS invariants $n_{0,i,j}$, $n_{1,i,j}$ and $n_{2,i,j}$  are given by 
\begin{center}
\begin{minipage}{14cm}
\underline{$n_{0,i,j}$ for trans.~III\ :} \qquad
\begin{tabular}{r|rrrrrrr}
i&j=0&j=1&j=2&j=3&j=4&j=5&j=6\\
\hline
0&0&  \bf{32}&     0&      0&      0&      0&      0\\
0&28&   32&     0&      0&      0&      0&      0\\
\hline
\end{tabular}
\vspace*{.4cm}

\underline{$n_{1,i,j}$ for trans.~III\ :}\qquad \begin{tabular}{r|rrrrrrr}
i&j=0&j=1&j=2&j=3&j=4&j=5&j=6\\
\hline
0&\bf{10}&   \bf{16}&     \bf{1}&      0&      0&      0&      0\\
1&310&  1408&   781&    16&     0&      0&      0\\
2&310&  4064&   6418&   1408&   10&     0&      0\\
\hline
\end{tabular}
\end{minipage}
\end{center}

\subsection{Example with $\hat E_7$ and $\hat E_6$ del Pezzo\label{E7example}}

In this appendix we present a compact Calabi-Yau example with an $\hat E_7$ and $\hat E_6$
del Pezzo which are in topological relation by intersecting non-trivially.
We also start with the quintic and add the point $\nu_6$ as in the following table.

\begin{center} 
\begin{tabular}{rlrrr|rr|rrr|rrr|rrrr}
&&&&& \multicolumn{2}{c|}{{quintic}} & \multicolumn{3}{c|}{{transition~I}} & 
\multicolumn{3}{c|}{{transition~II}} &   \multicolumn{4}{c}{{transition~III}} 
\\[.2cm]
\cline{6-17}
\rule[-0.2cm]{0cm}{0.8cm}
&&&&&&$l^{(1)}$&&$l_1^{(2)}$&$l_2^{(2)}$  &&$l_1^{(3)}$& $l_2^{(3)}$&&$l_1^{(4)}$& $l_2^{(4)}$& $l_3^{(4)}$     \\
&&&&&&-5&$x_0$                                 & -4 &-1 &$x_0$ &-3&-2 &$x_0$&-2&0&-1     \\
\hline 
&&&&&&&&&&&&&                        \\
$\nu_1$&=(\ -1,&0,&0,&0)& $x_1$& 1&  $u$      &2 &-1& $u_1$    &0& 1& $u_1$& 1& 0&-1                       \\
$\nu_2$&=(\ \ 0,&-1,&0,&0)&$x_2$&1&  $w_1$    &1 &0&  $u_2$   &0& 1& $v_1$& 0& 1&0                       \\ 
$\nu_3$&=(\ \ 0,&0,&-1,&0)&$x_3$&1&  $w_2$    &1 &0&  $v_1$   &1& 0& $v_2$&  1& -1&0                       \\ 
$\nu_4$&=(\ \ 0,&0,&0,&-1)&$x_4$&1&  $w_3$    &1 &0&  $u_3$   &0& 1& $s$ &   0& 1&0                       \\ 
$\nu_5$&=(\ \ 1,&1,&1, &1)&$x_5$&1&  $s$      &0 &1&  $v_2$   &1& 0& $u_2$&  0& 0&1                      \\ 
$\nu_6$&=(\ \ -2,&-1,&-1,&-1)& &&    $t$      &-1&1&          & &  & $t$ &  -1& 1&1                      \\
$\nu_7$&=(\ \ -1,&-1,&0,&-1)& & &             &  & &   $r$    &1&-1& $r$ &   1&-2&0     \\
\end{tabular}
\end{center}  

At $t=0$ we get an $\hat E_7$ del Pezzo. The topological data after the
transition I are $\chi=-164,\, h^{(1,1)}=2$ and
\begin{equation}
 [c_2]_1=42\ , \quad  [c_2]_2=50\ ,\quad
  \kappa_{111}=3,\ \ \kappa_{112}=\kappa_{122}= \kappa_{222}=5\ .
\end{equation} 
 In the first table summarizing the BPS 
 instantons we see $n_{1,0}=56$, which indicates the class with the $56$
lines of the $\hat E_7$ del Pezzo.
\begin{center}
\begin{tabular}{r|rrrrrrr}
i&j=0&j=1&j=2&j=3&j=4&j=5&j=6\\
\hline
0&0&    20&     0&      0&      0&      0&      0\\
1& \bf{56}&   2635&   5040&   190&    -40&    3&      0\\
\hline
\end{tabular}
\end{center}

After the transition II to the Calabi-Yau defined by $l^{(3)}_k$ we have 
an elliptic fibration with $\chi=-168,\, h^{(1,1)}=2$ and
\begin{equation}
 [c_2]_1=50\ , \quad  [c_2]_2=36,\quad
\kappa_{111}=\kappa_{112}=5, \qquad \kappa_{122}= 3\ .
\end{equation} 
We find the BPS number $n_{0,1}=18$ in the following table  
\begin{center}
\underline{$n_{i,j}$ for trans.~II\ :} \qquad
\begin{tabular}{r|rrrrrrr}
i&j=0&j=1&j=2&j=3&j=4&j=5&j=6\\
\hline
0&0&   \bf{18}&     -2&     0&      0&      0&      0\\
1&186&  2439&   442&    -512&   768&    -1024&  1280\\
\hline
\end{tabular}
\end{center}

In the following we will generate an $\hat E_7$ and $\hat E_6$ del Pezzo by considering 
both transitions simultaneously. After this transition III we find likewise an 
elliptically fibered Calabi-Yau space 
\begin{equation}
\chi=-148\ ,\quad  h^{(1,1)}=3\ , \quad [c_2]_1=78, \quad  [c_2]_2=36,\quad 
[c_2]_3=50\ , 
\end{equation} 
and the non-vanishing triple intersections      
\bea
\kappa_{111}&=& 21\ , \ \ 
\kappa_{112}=9\ , \ \
\kappa_{122}=\kappa_{223}=3\ , \ \
\kappa_{113}=18\ , \nn \\ 
\kappa_{123}&=&8\ , \ \ 
\kappa_{133}= 10\ ,\ \
\kappa_{233}=\kappa_{333}= 5\ . \ \
\eea
The BPS invariants $n_{0,i,j}$, $n_{1,i,j}$, $n_{2,i,j}$ and $n_{3,i,j}$   are
given by 
\begin{center}
\begin{minipage}{14cm}
\underline{$n_{0,i,j}$ for trans.~III\ :} \qquad
\begin{tabular}{r|rrrrr}
i&j=0&j=1&j=2&j=3&j=4\\
\hline
0&0&    16&     0&      0&      0 \\
1&0&     0&     0&      0&      0 \\
\hline
\end{tabular}
\vspace{.4cm}

\underline{$n_{1,i,j}$ for trans.~III\ :} \qquad
\begin{tabular}{r|rrrrr}
i&j=0&j=1&j=2&j=3&j=4\\
\hline
0&\bf{12}&   154&   16&      -2&      0\\
1&{\bf 10}&    {\bf 16}&  {\bf  1}&       0&      0\\
\hline
\end{tabular}
\vspace{.4cm}

\underline{$n_{2,i,j}$ for trans.~III\ :} \qquad
\begin{tabular}{r|rrrrr}
i&j=0&j=1&j=2&j=3&j=4\\
\hline
0& -2&  16&    154&   12&      0\\
1& \bf{32}&   2279&   4392&   130&   -32\\
2&-2& -32&  -20& 0& 0\\
\hline
\end{tabular}
\vspace{.4cm}

\underline{$n_{3,i,j}$ for trans.~III\ :} \qquad
\begin{tabular}{r|rrrrr}
i&j=0&j=1&j=2&j=3&j=4\\
\hline
0&0&    0&      16&     148&    16\\
1& -64& 528&    48158&  86440&  4291 \\
2& \bf{12}&  346&    2600&   1794&   128 \\
\hline
\end{tabular}
\end{minipage}
\end{center}

Here the $56$ of $E_7$ which appeared after transition I in the class $(1,0)$ is now 
splitted by the class $(1,1,0)$, i.e.~$n_{1,0,0}=12$, $n_{2,1,0}=32$ and
$n_{3,2,0}=12$. This corresponds to the splitting $SO(12)\times SU(2)\subset  E_7$ 
under which the representation $\bf{56}$ decomposes into 
$\bf{56}=(\bf{12,2})+ (\bf{32,1})$~\cite{Slansky}. In the class $(1,1,0)$ are
$n_{1,1,0}=10$ rational curves. As can be seen from the above table these seem to be part
of the ${\bf 27}={\bf 10}+{\bf 16}+{\bf 1}$ curves of an $\hat E_6$ del Pezzo
singularity which provides in this example the hidden singularity.

\section{Del Pezzo transitions $M^\mathfrak{g}_k \rightarrow M^\mathfrak{g}_{k+1}$} 
\label{mgn}

In section \ref{delPezzogeometries} we briefly discussed a del Pezzo transition 
starting with the hypersurface $\bbP(18|9,6,1,1,1)$. Such 
del Pezzo transitions are in fact ubiquitous in 
toric Calabi-Yau transitions. In particular, there are whole 
chains of transitions in which up to five $\hat E_6,\hat E_7$ or $\hat E_8$ del Pezzo surfaces 
can be blown up respectively. The corresponding  discrete family 
of $18$ reflexive polyhedra $\Delta^{\mathfrak{g}}_n$ in $4d$ is the 
the convex hull of the points 
\begin{equation} 
\label{trans}
\begin{array}{lrrr}
(\ -1,&0,&0,&0)\\
(\ \ 0,&-1,&0,&0)\\
(m^{\mathfrak{g}}_1,&m^{\mathfrak{g}}_2,&1,&0)\\
(m^{\mathfrak{g}}_1,&m^{\mathfrak{g}}_2,&0,&1)\\
(m^{\mathfrak{g}}_1,&m^{\mathfrak{g}}_2,&\nu_1^{(0)},&\nu_2^{(0)})\\
(m^{\mathfrak{g}}_1,&m^{\mathfrak{g}}_2,& \vdots,&\vdots)\\
(m^{\mathfrak{g}}_1,&m^{\mathfrak{g}}_2,&\nu_1^{(n)},&\nu_2^{(n)})
\end{array}
\quad {\rm with} \quad 
\begin{array}{rl} 
{\underline \nu}^{(0)}&=(-1,-1)\\
{\underline \nu}^{(1)}&=(0,-1)\\
{\underline \nu}^{(2)}&=(-1,0)\\
{\underline \nu}^{(3)}&=(1,1)\\
{\underline \nu}^{(4)}&=(1,-1)\\
{\underline \nu}^{(5)}&=(-1,1)\\
\end{array}
\quad {\rm and} \quad 
\begin{array}{rl} 
{\underline m}^{E_8}&=(3,2)\\
{\underline m}^{E_7}&=(2,1)\\
{\underline m}^{E_6}&=(1,1)\ .
\end{array}
\end{equation}  
For ${\rm rank}(\mathfrak{g})>5$ these $d=4$ polyhedra define hypersurfaces specified by
the anti-canonical bundle in the in toric variety and thus appear in the
list~\cite{kreuzer}. For ${\rm rank}(\mathfrak{g})<6$ one finds 
complete intersections associated to the nef partitions of toric 
varieties associated to reflexive polyhedra with $d>4$. 

The to \eqref{trans} corresponding compact Calabi-Yau manifolds are 
generically smooth elliptic fibrations over a del Pezzo base. 
In this fibration the worst degeneration of the fiber is of Kodaira type $I_1$ \cite{Kodaira}.
Let us denote the type of the elliptic fibration by the Lie algebra 
$\mathfrak{g}$ and the corresponding elliptic 
fibered Calabi-Yau over $\cB_n$ as $M^\mathfrak{g}_n$. As an collorary to 
the analysis of the elliptically fibered Calabi-Yau 
fourfolds \cite{klry}, one finds that the Euler number of 
the elliptic fibration $M^{\mathfrak{g}}_n$ is given by 
\begin{equation}
\chi(M^{\mathfrak{g}}_n)=- 2 h(\mathfrak{g}) \int_{\cB_n} c_1^2=2
h(\mathfrak{g}) (n-9),   
\end{equation}
here $h(\mathfrak{g})$ is the coxeter number of the associated Lie 
algebra. $h(\mathfrak{g})$ has been listed in Table \ref{tab-LocalSing}.  
For the smooth fibration spaces just described, 
$n+1$ K\"ahler classes of $M^\mathfrak{g}_n$ come from the del Pezzo 
base $\cB_n$ and $8-{\rm rank}(\mathfrak{g})$ come from the 
$8-{\rm rank}(\mathfrak{g})$ sections of the elliptic fiber. 
Using $\chi = 2 (h^{(1,1)}-h^{(2,1)})$ one infers
\bea \label{Hodge} 
h^{(1,1)}(M^\mathfrak{g}_n)&=&n+10-{\rm rank}(\mathfrak{g})\ ,\\
h^{(2,1)}(M^\mathfrak{g}_n)&=&h(\mathfrak{g})(9-n)+n+10-{\rm rank}(\mathfrak{g})\ . \nn
\eea
Many of the $M^\mathfrak{g}_n$ have already appeared in the physics 
literature as, for example,~in refs.~\cite{Morrison:1996pp,LSTY,Klemm:1996hh}. In section \ref{delPezzogeometries} we discussed the
transition $M_0^{E_8} \ \rightarrow \ M_1^{E_8}$.


\end{document}